\documentclass[%
 aip,
 jmp,%
 amsmath,amssymb,
 reprint,%
author-year,%
]{revtex4-2}
\usepackage{lineno}
\usepackage{graphicx}
\usepackage{dcolumn}
\usepackage{bm}
\usepackage[]{float}
\usepackage{amsmath}
\usepackage{hyperref}
\usepackage{amssymb}
\usepackage{mathtools}
\usepackage{comment}
\usepackage{amsmath}
\usepackage{algpseudocode}
\usepackage{algorithm}
\usepackage[table]{xcolor}
\usepackage{xcolor}
\usepackage{soul}

\definecolor{darkergreen}{RGB}{0,128,0}
\definecolor{revcolor}{rgb}{1.0, 0.0, 0.0}

\def\marg{1ex}

\newcommand{\popI}{\mathcal{P}_1}
\newcommand{\popII}{\mathcal{P}_2}
\newcommand{\NI}{\mathcal{N}_1}
\newcommand{\NII}{\mathcal{N}_2}

\newcommand{\PII}{\text{P}_2}
\newcommand{\T}{\mathcal{T}}
\newcommand{\rh}{\nu_{h}}
\newcommand{\rl}{\nu_{\ell}}
\newcommand{\rhI}{\nu_{h\text{,1}}}
\newcommand{\rlI}{\nu_{\ell,1}}
\newcommand{\rhII}{\nu_{h\text{,2}}}
\newcommand{\rlII}{\nu_{\ell,2}}
\newcommand{\rt}{\nu_\text{t}}
\newcommand{\rtI}{\nu_\text{t,1}}
\newcommand{\rtII}{\nu_\text{t,2}}
\newcommand{\C}{\mathcal{C}}
\newcommand{\Wb}{\mathcal{W}_\text{b}}
\newcommand{\Wc}{\mathcal{W}_\text{s}}
\newcommand{\SII}{\mathcal{S}_\text{c}}
\newcommand{\Sb}{\mathcal{S}_\text{b}}
\newcommand{\varSb}{\sigma^{2}_\text{b}}
\newcommand{\varSc}{\sigma^{2}_\text{c}}

\newcommand{\sps}{\,\text{spikes/s}}

\begin{document}

\preprint{AIP/123-QED}

\title{A theoretical framework for learning through structural plasticity}

\author{Gianmarco Tiddia}
\email{gianmarco.tiddia@dsf.unica.it}
\affiliation{Department of Physics, University of Cagliari, Italy}
\affiliation{Istituto Nazionale di Fisica Nucleare (INFN), Sezione di Cagliari, Cagliari, Italy}
\author{Luca Sergi}%
\affiliation{Department of Physics, University of Cagliari, Italy}
\affiliation{Istituto Nazionale di Fisica Nucleare (INFN), Sezione di Cagliari, Cagliari, Italy}

\author{Bruno Golosio}
\affiliation{Department of Physics, University of Cagliari, Italy}
\affiliation{Istituto Nazionale di Fisica Nucleare (INFN), Sezione di Cagliari, Cagliari, Italy}

\date{\today}

\begin{abstract}
A growing body of research indicates that structural plasticity mechanisms are crucial for learning and memory consolidation.
Starting from a simple phenomenological model, we exploit a mean-field approach to develop a theoretical framework of learning through this kind of plasticity, capable of taking into account several features of the connectivity and pattern of activity of biological neural networks, including probability distributions of neuron firing rates, selectivity of the responses of single neurons to multiple stimuli, probabilistic connection rules and noisy stimuli. More importantly, it describes the effects of stabilization, pruning and reorganization of synaptic connections. This framework is used to compute the values of some relevant quantities used to characterize the learning and memory capabilities of the neuronal network in training and testing procedures as the number of training patterns and other model parameters vary. The results are then compared with those obtained through simulations with firing-rate-based neuronal network models.
\end{abstract}

\maketitle

\section{\label{sec:introduction} Introduction}
Together with temporary and reversible changes of synaptic efficacy such as short and long-term plasticity mechanisms, structural changes in the synaptic morphology of the network are fundamental mechanisms that take place in healthy brains. These changes occur at longer time scales than the short or long-term mechanisms mentioned above and consist in the stabilization, creation of new synapses, or erasure of synapses that have not been stabilized \citep{Lamprecht2004, Tetzlaff2012}. This type of synaptic plasticity, called \textit{structural plasticity}, can be spontaneous but also experience-based \citep{Butz2009}, and it has a key role in the stabilization of new concepts that need to be kept in memory after learning \citep{Fu2011}.\\
Indeed, it is known that neurotransmitters can be neurotrophic factors, i.e. participate in the growth or suppression of dendritic spines, synapses, axons, and dendrites \citep{Mattson1988, Lamprecht2004, Richards2005}. Thus, structural plasticity is a neural-activity-driven mechanism, which can increase or decrease the number of synapses. Such modifications are flanked by a homeostatic kind of structural plasticity, which has a balancing effect achieved by adding or removing synapses, as described in \citep{Fauth2016}.\\
Moreover, the number of synapses in the brain can change over time. In \cite{Huttenlocher1979} it is shown that synaptic density in the human cortex reaches the highest values at $1$-$2$ years age, it drops during adolescence and stabilizes between age $16$-$72$, followed by a slight decline. However, although synaptic density remains approximately stable during adulthood, rewiring of network connections occurs as well in order to efficiently store new memories \citep{Navlakha2015, Zito2002}. The activity-dependent connectivity changes, together with the rearrangement of synapses lead to a fine-tuning of the brain's circuits \citep{sakai2020}. Indeed, some synapses can be strengthened through long-term potentiation (LTP) and new connections can be formed next to the already potentiated ones to further enhance synaptic transmission. On the other hand,
when the presynaptic and postsynaptic neuron activities have a low correlation, their connection is more likely to be removed. The latter process is called synaptic pruning and it is considered essential for optimizing activity propagation and memory capacity \citep{Chklovskii2004, Knoblauch2014, Knoblauch2016}.
Furthermore, it is commonly believed that synaptic pruning and rewiring dysfunction are neural correlates of developmental disorders such as autism or schizophrenia \citep{Bourgeron2009, Moyer2015}, leading to, respectively, a higher or lower synaptic density with respect to neurotypical subjects \citep{Hutsler2010, Pagani2021, Glantz2000}.\\
In the last decades, computational neuroscience has 
investigated brain dynamics at different scales, from cellular \citep{Markram2015} to mesoscopic and macroscopic through mean-field approaches \citep{Wilson1972, Amit1997, Hopfield1984, Renart2004, SanzLeon2013, diSanto2018, Capone2019, Carlu2020}. 
Regarding synaptic plasticity, computational models were 
mostly focused on plasticity mechanisms that involve strengthening or weakening of existing synapses, like short-term plasticity (STP) \citep{tsodyks1998} or spike-timing-dependent plasticity (STDP) \citep{Gutig2003} and on their role in short-term, long-term, working memory and learning \citep{Mongillo2008, Tiddia2022_WM, Song2000, Bi2001, ThaCo, Capone2022}. Only in recent times, computational models of structural plasticity and connectivity rearrangements during learning were developed, showing intriguing results. \cite{Knoblauch2014} and \cite{Knoblauch2016} describe a model of structural plasticity based on "effectual connectivity", defined in these works as the fraction of synapses able to represent a memory stored in a network. By structural plasticity, effectual connectivity is improved, since synapses that do not code for the memory are moved in order to optimize network's connectivity. Their model defines synapses using a Markov model of three states: potential (i.e. not instantiated), instantiated but silent or instantiated and stabilized. Structural plasticity is thus related to the passage of the synapses from a potential state to an instantiated state (and vice versa), whereas changes only related to the synaptic weight are described by the potentiation of the instantiated synapses. With such a model, it is possible to show that networks with structural plasticity have higher or comparable memory capacity to networks with dense connectivity and it is possible to explain some cognitive mechanisms such as the spacing effect \citep{Knoblauch2014}.\\
\cite{Spiess2016} simulated a spiking neural network with structural plasticity and STDP, showing that structural plasticity reduces the amount of noise of the network after a learning process, thus making the network able to have a clearer output. Furthermore, such a network with structural plasticity shows higher learning speed than the same network with only STDP implemented.\\
Some new insights about the importance of synaptic pruning are also shown in \cite{Navlakha2015}, in which different pruning rates were studied suggesting that a slowly decreasing rate of pruning over time leads to more efficient network architectures.\\
The model proposed in this work considers two populations of neurons, $\popI$ and $\popII$, with synaptic connections directed from the first to the second population (i.e., in a feed-forward fashion).
During the training phase, $\popI$ receives an input stimulus, while $\popII$ receives a contextual stimulus.
The model assumes that synaptic stabilization is a probabilistic process driven by pre- and postsynaptic spiking activity. For each pattern given in input to the model, only a small fraction of neurons contribute to this process.
Connection stabilization is complemented by a rewiring process: connection pruning, which eliminates unstabilized connections, and creation of new connections, which restores network balance.
This framework does not specifically refer to a particular region of the brain; rather, it is built on characteristics that are ubiquitous for several brain areas.\\
As discussed above, the biochemical and biophysical mechanisms underlying structural plasticity are extremely complex and only partially understood to date. For this reason, rather than attempting to build a biologically detailed model, this work exploits a relatively simple phenomenological model, including both the activity-driven and the homeostatic contributions; despite the lower complexity, this model accounts for the effects of structural plasticity in terms of the stabilization of synaptic connections between neurons with a high activity correlation as well as those of pruning and rewiring the connections for which this correlation is lower. This approach is also justified by the requirement for a simple and effective computational model suitable for simulating networks with a relatively large number of neurons and connections and for representing learning processes with sizable numbers of training and testing patterns.
The model will then serve as the foundation for the creation of a mean-field-based theoretical framework for learning through synaptic plasticity capable of accounting for a variety of biological network properties. This framework is used in a training and testing procedure to characterize learning and memory capacity of plastic neuronal networks as the number of training patterns and other model parameters vary. The results are then compared with those obtained through simulations based on feed-forward firing-rate-based neuronal networks.\\
The proposed approach is capable of accounting for different probabilistic connection rules, firing rate probability distributions, presence of noise in stimuli, thus providing a general framework to study the impact of structural plasticity on learning in large-scale neuronal network models.

\section{\label{sec:model_description} Model description}
This section describes the model proposed in this work. A pseudo-code of the algorithm used to model the structural plasticity mechanism is provided in Algorithm \ref{alg:stabilization}. The neuronal network consists of two neuron populations, $\popI$ and $\popII$, with $10^5$ neurons each. The exchange of information between the two populations takes place through the feed-forward connections from the population $\popI$ to the population $\popII$, which in the model are on average $5\cdot 10^3$ per neuron of $\popII$ for a total of $5\cdot 10^8 $ connections. Each connection has an initial baseline synaptic weight $\Wb$ (\textit{initialization} in Algorithm \ref{alg:stabilization}).
To mimic the activity of $\popI$ in response to an external input signal (e.g., a visual input), the model assigns to each neuron of this population a value of firing rate derived from a predefined firing rate distribution. This way, an input is modeled as a firing rate pattern of $\popI$.
During the training stage, using the same approach previously described for $\popI$, the population $\popII$ is injected with a stimulus (e.g., auditory), that we identify as a \textit{contextual stimulus} (as proposed in \cite{ThaCo}). In this phase we assume that the activity of $\popII$ is entirely derived by the contextual stimulus, neglecting the contribution of the connectivity between $\popI$ and $\popII$ (\textit{training} in Algorithm \ref{alg:stabilization}). 

The structural plasticity model follows the categories proposed by \cite{Fauth2016}, i.e., activity-dependent and homeostatic. The firing rate patterns of the two populations have a role in the activity-dependent structural plasticity.
The potentiation and stabilization of a synaptic connection occurs when the firing rates of both the presynaptic and the postsynaptic neurons are concurrently above a certain threshold (the definition of which varies depending on the firing rate distribution). In our model, the synaptic weight of a stabilized connection increases from $\Wb$ to a value $\Wc > \Wb$ (\textit{synapses potentiation and stabilization} in Algorithm \ref{alg:stabilization}). This is a computationally effective way of taking into account the several biological mechanisms that concur in the stabilization of the connection between two neurons.  
We flank this mechanism with synaptic rewiring, which consists of the mechanism of pruning of the connections that have not been stabilized yet together with the creation of new connections handled by homeostatic structural plasticity (\textit{synapse pruning} and \textit{synapse creation} in Algorithm \ref{alg:stabilization}). We apply synaptic rewiring on the simulation periodically after a certain number of simulation steps. Once a synaptic connection has been stabilized, it will maintain the synaptic weight $\Wc$, without the possibility of returning to the initial state $\Wb$. Thus, these connections are prevented from being pruned in further simulation steps.\\
The sets of input and contextual patterns used for network training are independent firing-rate patterns of the two populations randomly generated from predefined firing-rate probability distributions. The training process is performed using $\T$ independent input patterns, together with the corresponding contextual stimuli. 

During training, when both input and contextual stimulus are used, a fraction of the neurons of the population $\popII$ assumes a high value of firing rate (i.e., above threshold), becoming thus representative of that input. These neurons, called {\it coding}, or {\it selective} neurons, play a vital role in input coding. The average input signal to these neurons will be called $\langle \SII \rangle$. The non-selective neurons of $\popII$ will instead be called {\it non-coding} or {\it background} neurons, and their average input signal will be indicated with $\langle \Sb \rangle$. The average incoming signals to background and coding neurons of $\popII$ (i.e., $\langle \Sb \rangle$ and $\langle \SII \rangle$) is evaluated in the test phase, during which an input pattern is provided to $\popI$ and the signal incoming to neurons of $\popII$ is entirely derived from the connectivity between the two populations optimized during training thanks to the structural plasticity mechanism. Input patterns in the test phase are derived from the training patterns with the addition of noise.
The proposed model accounts for the ability of the network to learn the association between input patterns and the corresponding contextual stimuli.

\begin{figure}[H]
\centering
\includegraphics[width=0.95\columnwidth]{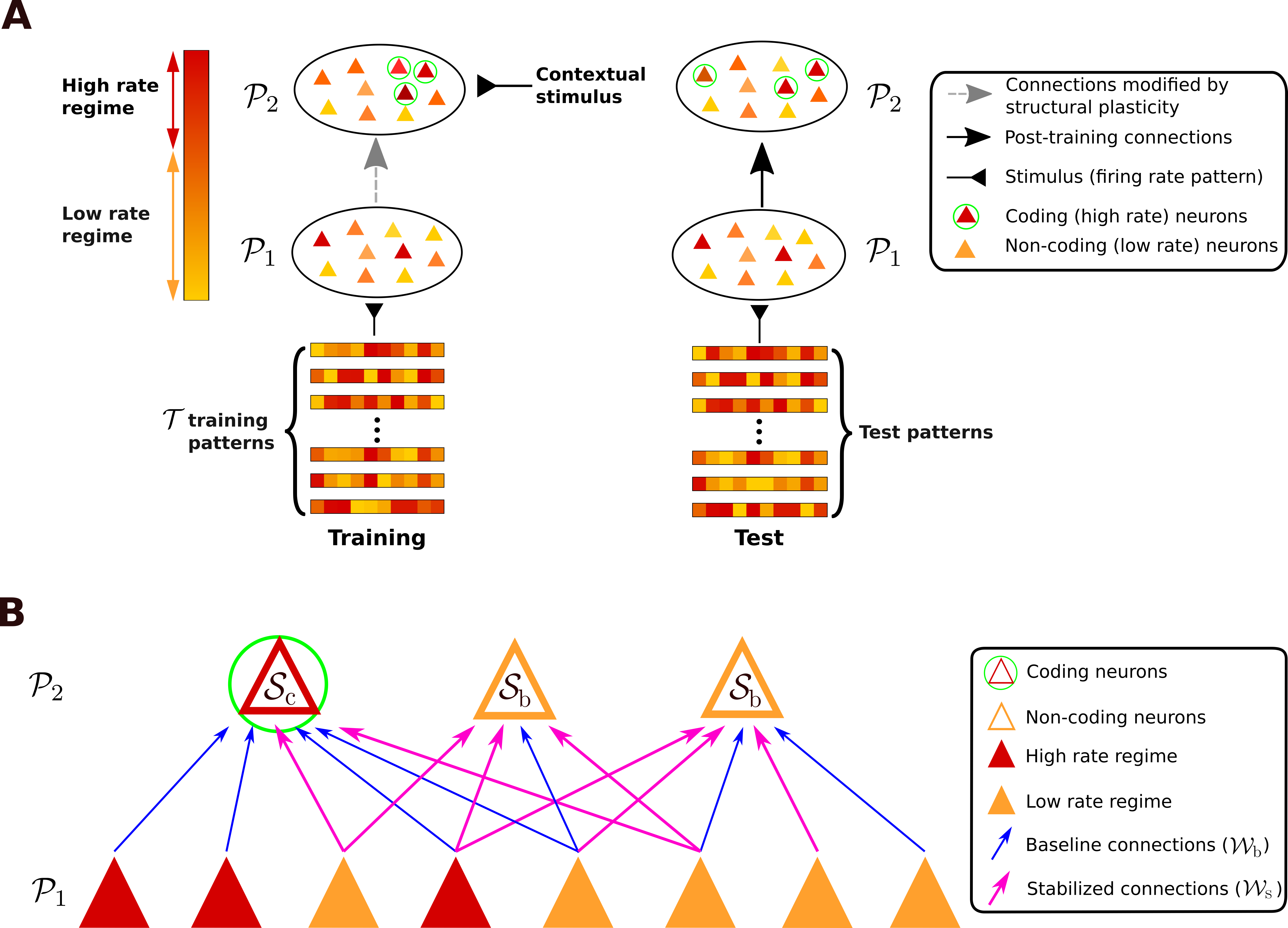}
\caption{Schematic representation of the network model. The triangles represent neurons of the network, with their rate indicated by the color (from yellow to red). \textbf{(A)} During training (left), a stimulus is injected into $\popI$, and a contextual stimulus is injected into $\popII$. In this phase, structural plasticity occurs by refining and reorganizing the connectivity between the two neuron populations (dashed arrow). Moreover, during training we assume that the activity of $\popII$ is only derived from the contextual stimulus, neglecting the contribution of the connections between the populations. The stimuli are randomly generated from a predefined firing rate probability distribution.
In the test phase (right), a pattern is injected without the corresponding contextual stimulus and the connections between the two populations project the input to the neurons of $\popII$. The test patterns are derived from the training patterns with the addition of noise.
\textbf{(B)} Detail of the test phase, in which is shown a subset of neurons of $\popI$ and $\popII$ (triangles) connected through connections (arrows). The triangles are colored so that amber triangles fall in the low rate regime and red ones in the high rate regime, and the connections are distinguished between baseline (blue, with synaptic weight $\Wb$) and stabilized (fuchsia, with synaptic weight $\Wc$). The neuron of $\popII$ on the left is a coding neuron, i.e., it is representative of the input injected, and thus receives an input signal $\SII$. The other neurons of $\popII$, being non-coding neurons, receive a background signal $\Sb$. 
\label{fig:net_scheme}}
\end{figure}

A diagram of the training and testing processes is shown in Figure \ref{fig:net_scheme}. 

\setlength{\intextsep}{10mm}
\begin{algorithm}[H]
\caption{Pseudo-code of structural plasticity during training}
\label{alg:stabilization}
\begin{algorithmic}
\State
\State \textbf{\color{blue}{ INITIALIZATION}}
\State \textcolor{darkergreen}{\# creating initial connectivity}
\For{$i$ from 1 to $\NII$}
    \State $\C_i$ = extractNumberOfConnections($\C$, connection\_rule)
    \For{$c$ from 1 to $\C_i $}
        \State \textcolor{darkergreen}{\# presynaptic neuron index is extracted randomly from $\popI$}
        \State $j$ = generateRandomInteger(1, $\NI$)
        \State createConnection($i,j$)
        \State $\mathcal{W}_{i,j} = \Wb$
    \EndFor
\EndFor
\State
\State
\State \textbf{\color{blue}{ TRAINING}}
\State \textcolor{darkergreen}{\# training on $\T$ patterns}
\For{$t$ from 1 to $\T$}
    \State \textcolor{darkergreen}{\# providing to the network a firing rate pattern}
    \State stimulusInjection()
    \For{$i$ from 1 to $\NII$}
        \For{$c$ from 1 to $\C_i$}
            \State \textcolor{darkergreen}{\# retrieve index of the presynaptic neuron for connection $c$}
            \State $j$ = sourceNodeIndex($i, c$)
            \State \textbf{\color{blue}
            \State
            \State{SYNAPSES POTENTIATION AND STABILIZATION}}
            \State
            \textcolor{darkergreen}{\# synapse potentiation and stabilization if both neurons have high rate}
            \If{$\nu_i \geq \nu_{th}$ and $\nu_j \geq \nu_{th}$} 
                \State $\mathcal{W}_{i,j} = \Wc$
            \EndIf
            \State
            \State \textbf{\color{blue}{SYNAPSES PRUNING}}
            \State \textcolor{darkergreen}{\# connection pruning after every rewiring\_step}
            \If{isMultipleInteger($t$, rewiring\_step) }
                \If{$\mathcal{W}_{i,j} = \Wb$}
                    \State pruneConnection($i,j$)
                \EndIf
            \EndIf
        \EndFor
        \State
        \State \textbf{\color{blue}{SYNAPSES CREATION}}
        \State \textcolor{darkergreen}{\# creation of new connections after every rewiring step}
        \If{isMultipleInteger($t$, rewiring\_step)}
            \State $k$ = numberOfStabilizedConnections($i$)
            \State $\C_i$ = extractNumberOfConnections($\C$, conn\_rule)
            \For{$c$ from $k$ to $\C_i$}
                \State $j$ = chooseRandomNeuronFromPopulation($\popI$)
                \State createConnection($i,j$)
                \State $\mathcal{W}_{i,j} = \Wb$
            \EndFor
        \EndIf
    \EndFor
\EndFor
\end{algorithmic}
\end{algorithm}

The biological motivations for the choices we adopted for the model design are reported in the following section.

\subsection{Biological justification of the model choices}
The model described above has several biologically motivated features. First, the average number of connections per neuron has been chosen in agreement with experimental estimation \citep{Pakkenberg2003}. Training and test patterns are generated so that a fraction of the neurons targeted by the external stimuli show a high-rate regime in response to the input.

During training, when input and contextual stimuli are used, a fraction of the neurons of $\popII$ become representative of that input. Indeed, the existence of neurons showing selective firing rates in response to specific stimuli is largely confirmed by experimental results. Additionally, the neuron populations assume a continuous firing rate distribution when targeted by an external stimulus, with the firing rate of each neuron derived by the distribution. In agreement with the experimental observations, we employ a lognormal distribution of the firing rate \citep{Roxin2011}. 

Regarding the structural plasticity mechanism, we followed the description of \cite{Fauth2016}, thus dividing the structural mechanism into two categories: activity-dependent and homeostatic. Activity-dependent mechanisms lead to stabilization and potentiation of a synapse when the pre- and postsynaptic neurons assume a high-rate regime. Indeed, it is believed that the potentiation of the synapses that connect neurons with strongly correlated activity is accompanied by greater stability over time \citep{Yang2008}. As a result, the synapse increases the synaptic efficacy to a value $\Wc$ than the baseline value $\Wb$. This can be seen as an effect of LTP, according to which we have the formation of new synapses, but also the increase of synaptic efficacy of the existing ones. Thus, the mechanism of stabilization proposed here embraces both structural and functional synaptic modifications, both mediated by LTP \citep{Muller2002}.

Homeostatic effects of structural plasticity take into account the mechanisms of synaptic pruning of the connections that have not been stabilized and the creation of novel synaptic contacts. We modeled these effects through synaptic rewiring, according to which non-stabilized connections are periodically pruned and new connections are created. This mechanism is not performed after every simulation step since it is known that the homeostatic effects take place on a longer time scale than the activity-dependent ones \citep{Fauth2016}. The mechanism of synaptic rewiring will be discussed in detail later in this work. Stabilized connections cannot be affected, by definition, by synaptic pruning, and will keep their synaptic weight to $\Wc$, without the possibility of returning to the initial state $\Wb$. This approach is a simplified way of representing the biological structural changes that make a stabilized connection strong and durable.\\
In the next section, we derive the mean-field equations of the model, whose description and parameters are summarized in the Tables in Appendix \ref{app:tables}.

\section{\label{sec:rate-cont} Theoretical Derivations}
Here we introduce the theoretical framework of the model. As mentioned earlier, in this model the injection of an external input to a neuron population is represented by neurons showing firing rate patterns generated from a probability distribution $\rho(\nu)$. The distinction between high-rate and low-rate neurons is based on two rate thresholds, $\rtI$ for the population $\popI$ and $\rtII$ for the population $\popII$. The value of these thresholds is related to the fraction of neurons above the threshold (i.e., high-rate) for the two populations, $\alpha_1$ and $\alpha_2$, respectively, by the equations:
\begin{equation}
    \begin{split}
    \alpha_1 &= \int_{\rtI}^{\infty} \rho (\nu) d\nu\\
    \alpha_2 &= \int_{\rtII}^{\infty} \rho (\nu) d\nu .\\
    \end{split}
\end{equation}
Thus, the average number of high-rate neurons when an external stimulus is provided is
\begin{equation}
\label{eq:Nh}
\begin{split}
     N_{h,1}&=\NI \alpha_1 \\
     N_{h,2}&=\NII \alpha_2,
     \end{split}
\end{equation}
where $\NI$ and $\NII$ are the number of neurons of $\popI$ and $\popII$, respectively. The average rates of the neurons of $\popI$ below and above threshold, $\langle \rlI \rangle$ and $\langle \rhI \rangle$, can be computed from the firing rate distribution as:
\begin{equation}
    \label{eq:av_rate_cont}
    \begin{split}
        \langle \rlI \rangle &= \int_{0}^{\rtI} \nu \rho (\nu) d\nu \Big/ \int_{0}^{\rtI} \rho (\nu) d\nu = \frac{1}{\beta_1}\int_{0}^{\rtI} \nu \rho (\nu) d\nu\\
        \langle \rhI \rangle &= \int_{\rtI}^{\infty} \nu \rho (\nu) d\nu \Big/ \int_{\rtI}^{\infty} \rho (\nu) d\nu = \frac{1}{\alpha_1} \int_{\rtI}^{\infty} \nu \rho (\nu) d\nu ,
    \end{split}
\end{equation}
where $\beta_1 = 1- \alpha_1$. Similar equations can be used to compute
$\langle \rlII \rangle$ and $\langle \rhII \rangle$. In general, the thresholds $\rtI$ and $\rtII$ can be different, however, in this work, we assume that the thresholds are the same. From these equations, the average firing rate can be expressed as
\begin{equation}
    \langle \nu \rangle= 
    \int_{0}^{\infty} \nu \rho (\nu) d\nu
    = \beta_1 \langle \rlI \rangle + \alpha_1 \langle \rhI \rangle 
    = \beta_2 \langle \rlII \rangle + \alpha_2 \langle \rhII \rangle .
\end{equation}

The theoretical framework aims to provide an estimation of how well such a feed-forward network is able, after a training process, to store the learned patterns only through the connections changed by the structural plasticity mechanism. As a result of training, a subset of neurons of $\popII$, called \textit{coding neurons}, should receive during the test phase a higher input than the rest of the neurons of $\popII$, which we call \textit{background neurons}. In such a framework, an effective approach for estimating the capacity of the network to recognize the pattern would be to evaluate, during the test phase, the signal-difference-to-noise-ratio (SDNR) using the formula
\begin{equation}
\label{eq:SDNR}
    \text{SDNR} = \dfrac{|\langle \SII \rangle - \langle \Sb \rangle |}{\sigma_b},
\end{equation}
where $\langle \SII \rangle$ and $\langle \Sb \rangle$ are, respectively, the average input signal to coding and background neurons of the $\popII$ population due to the connections coming from the $\popI$ population and $\sigma_b$ is the standard deviation of the signal received by the background neurons.
This choice is justified by the need to evaluate the memory capacity associated with the plasticity of the connections from $\popI$ to $\popII$. In Appendix \ref{app:sdnr_memory_capacity} we derive the relation between SDNR and the probability of correctly recalling a pattern after training:

\begin{equation}
\label{eq:sdnr_probability_correct_recall}
    P_\text{C} \simeq  \dfrac{1}{2}\Bigr[1 + \text{erf}\Bigl(\frac{\text{SDNR}}{\sqrt{8}}\Bigl)\Bigr].
\end{equation}

This equation can be used to set a lower limit to the SDNR ensuring a probability of correct recall greater than a predetermined threshold $P_{\text{C}}$ (e.g., $P_\text{C} = 0.95$). Since the SDNR depends on the number of training patterns $\T$, it can be used to estimate the highest number of patterns viable for training which ensures retrieval during testing with a probability $P_\text{C}$ (i.e., $\T_{\text{max}}$).
Training the network with an excessive number of patterns surpassing this maximum value makes the SDNR insufficient for distinguishing coding from background signals, increasing the likelihood of incorrect pattern recall.

During training, when input and contextual stimuli are provided to the network, a connection is stabilized if both the presynaptic and the postsynaptic neurons assume a firing rate above the threshold.\\
In this work, we use a lognormal distribution of the firing rates for the continuous model.
Indeed, it is known that rate distribution in the cortex is long-tailed and skewed with a lognormal shape 
\citep{Roxin2011}. More details on the implementation of the lognormal distribution and the choice of the average low and high rates can be found in Appendix \ref{app:distr_fr}.
Nevertheless, the following derivations are valid for a generic probability distribution $\rho (\nu)$.

The test set consists of $V$ firing-rate patterns of the neurons of $\popI$, randomly extracted from the $\T$ input patterns of the train set.
Here we consider the case where the patterns are unchanged, thus each input pattern of the test set is identical to an input pattern of the train set. In a later section, we will discuss the effect of altering these patterns by adding noise.
To estimate the values of $\langle\SII\rangle$, $\langle\Sb\rangle$, and $\varSb$ we can start by computing the average number of incoming stabilized connections for every neuron of $\popII$ after a training process in which $\T$ independent patterns are given. This is useful since the input to a neuron of $\popII$ driven by the connections from $\popI$ is determined by the product between the synaptic weights of the connections and the rate of the respective presynaptic neurons.\\
The probability that a generic connection is stabilized in a single example is $\alpha_1 \alpha_2$, and thus the probability that it is not stabilized after $\T$ training examples is
$(1-\alpha_1\alpha_2)^{\T}$.
The probability $p_{\T}$ that a connection is stabilized in at least one of the $\T$ training examples is given by the complement of the previous expression:
\begin{equation}
\label{eq:p}
     p_{\T}= 1-(1-\alpha_1 \alpha_2)^{\T}.
\end{equation}
Being $k$ the number of stabilized connections for a neuron, the average number of stabilized connections per neuron of $\popII$ is defined as the product of the expression above by the number of incoming connections per neuron $\C$. This way, the average number of stabilized connections per neuron is:
\begin{equation}
\label{eq:av_k}
     \langle k \rangle=\C \Bigl[ 1 - (1-\alpha_1\alpha_2)^{\T} \Bigr] = \C p_{\T}.
\end{equation}
For each neuron we have on average $\langle k \rangle$ stabilized connections with synaptic weight $\Wc$ and $\C- \langle k \rangle$ non-stabilized connections with the baseline synaptic weight $\Wb$. For simplicity, the previous equation and the following derivations assume that the number of incoming connections per neuron is constant. However, in the most general case, this number can be driven from a distribution. In this regard, Section \ref{subsec:poiss-indegree} describes the changes in the theoretical framework in the case of Poisson-distributed in-degrees.\\
First of all, we calculate the input signal to a background neuron of $\popII$.
Let $\C$ be the number of incoming connections to this neuron, $k$ the number of stabilized connections, $\nu_1, \dotsi, \nu_k$ the firing rates of the presynaptic neurons of the stabilized connections, and $\xi_1, \dotsi, \xi_{\C - k}$ the firing rates of the presynaptic neurons of the non-stabilized connections. The input signal is then
\begin{equation}
    \label{eq:sb_single_neuron}
    \Sb = \Wc \sum_{i=1}^{k}\nu_i + \Wb \sum_{i=1}^{\C - k} \xi_i.    
\end{equation}
To calculate the average background signal we should average the expression given by Equation \eqref{eq:sb_single_neuron} over all the possible values of $k$ and of the firing rates. Being $P(k)$ the probability that $k$ of these connections are stabilized, the probability of having $k$ stabilized connections and rates in the range $(\nu_1, \nu_1+d\nu_1), \dotsi, (\nu_k, \nu_k + d\nu_k)$, $(\xi_1, \xi_1+d\xi_1), \dotsi, (\xi_{\C-k}, \xi_{\C-k} + d\xi_{\C-k})$ is $P(k)\rho (\nu_1)\dotsi \rho (\nu_k) \rho (\xi_1) \dotsi \rho (\xi_{\C-k}) d\nu_1 \dotsi d\nu_k d\xi_1 \dotsi d\xi_{\C-k}$. Thus
\begin{equation}
\label{eq:av_sb_cont}
\begin{split}
    \langle \Sb \rangle &= \sum_{k} P(k) \int d\nu_1 \dotsi \int d\nu_k \int d\xi_1 \dotsi \int d\xi_{\C - k} \rho (\nu_1) \dotsi \rho (\nu_k) \rho (\xi_1) \dotsi \rho (\xi_{\C - k}) \cdot \\
    &\cdot\Bigl[ \Wc (\nu_1 + \dotsi + \nu_k) + \Wb (\xi_1 + \dotsi + \xi_{\C-k})\Bigr] = \\
    &= \sum_{k} P(k) \Bigl[ \Wc k \langle \nu \rangle + \Wb (\C-k) \langle \nu \rangle \Bigr]
    = [ \Wc \langle k \rangle 
    +  \Wb (\C - \langle k \rangle) ] \langle \nu \rangle ,
\end{split}
\end{equation}

where we used the fact that $\int \nu \rho(\nu) d\nu = \int \xi \rho(\xi) d\xi = \langle \nu \rangle$. In this equation, we can clearly observe two distinct contributions: one related to stabilized connections, which depends on the mean value $\langle k \rangle$, and the other related to baseline connections, which depends on $\C - \langle k \rangle$. Both these contributions are multiplied by the average firing rate of $\popI$ neurons, which is related to the non-selectivity of the incoming signal.\\
The variance of the background signal can be derived by applying its definition:
\begin{equation}
\label{eq:var_sb_cont_preliminary}
\begin{split}
    \varSb &=\langle(\Sb - \langle \Sb \rangle)^2\rangle = \sum_k P(k) \int d\nu_1  \dotsi \int d\nu_k \int d\xi_1 \dotsi \int d\xi_{\C-k} \rho(\nu_1) \dotsi \rho(\xi_{\C-k}) \cdot \\
    &\cdot \Bigl[ \Wc \sum_{i=1}^{k}\nu_i
    +\Wb \sum_{i=1}^{\C-k} \xi_i 
    - [ \Wc \langle k \rangle 
    +  \Wb (\C - \langle k \rangle) ] \langle \nu \rangle \Bigr]^2.
\end{split}
\end{equation}
Taking advantage of the equality $\langle k \rangle=k+(\langle k \rangle-k)$, we can rewrite:
\begin{equation}
   \Wc \langle k \rangle
    + \Wb (\C - \langle k \rangle)
    = \Wc k + \Wc (\langle k \rangle - k) + \Wb \Bigl[(\C-k) + (k-\langle k \rangle ) \Bigr] =
\end{equation}
\begin{equation*}
     =\Wc k + \Wb (\C- k) + \Wc (\langle k \rangle - k)  + \Wb (k - \langle k \rangle) .
\end{equation*}
Inserting this last expression in Equation \eqref{eq:var_sb_cont_preliminary} and rewriting the terms with the multiplicative factors $k$ and $\C - k$ with summations, such as for example $\Wc k \langle \nu \rangle = \Wc \sum_{i=1}^{k} \langle \nu \rangle$, we obtain:
\begin{equation}
\label{eq:var_sb_cont}
\begin{split}
    \varSb &= \sum_k P(k) \int d\nu_1 \dotsi \int d\xi_{\C-k} \rho(\nu_1) \dotsi \rho(\xi_{\C-k})
    \Bigl[ \Wc \sum_{i=1}^k (\nu_i - \langle \nu \rangle) +\Wb \sum_{i=1}^ {\C-k} (\xi_i - \langle \nu \rangle) + \\ 
    & + (k-\langle k \rangle)(\Wc-\Wb)\langle \nu \rangle \Bigr]^2 .
\end{split}
\end{equation}
The mixed terms of the equation above are null because $\int \rho (x) (x-\langle x \rangle)dx = 0$, ergo we can write the variance of the background signal as follows:
\begin{equation}
\label{eq:var_sb2_cont}
\begin{split}
    \varSb &= \sum_k P(k) \int d\nu_1 \dotsi \int d\xi_{\C-k} \rho(\nu_1) \dotsi \rho(\xi_{\C-k}) \Bigl[ \Wc^2 k \langle (\nu - \langle \nu \rangle)^2 \rangle + \Wb^2 (\C - k) \langle (\nu - \langle \nu \rangle)^2 \rangle +\\
    & + (\Wc - \Wb)^2 (k - \langle k \rangle)^2 \langle \nu \rangle ^2 \Bigr] = \Bigl[ \Wc^2 \langle k \rangle + \Wb^2 (\C - \langle k \rangle)\Bigr] \sigma_{\nu}^2 + (\Wc - \Wb)^2 \sigma_{k}^2 \langle \nu \rangle ^2 ,
\end{split}
\end{equation}
where $\sigma^{2}_{k}=\langle (k - \langle k \rangle) \rangle$. In the previous formula, we note two contributions depending respectively on the variance of the firing rate and the variance of the number of stabilized connections. The value of the variance of $k$ is not shown here, but is derived in Appendix \ref{app:var_k}, whereas the variance of the rate is, by definition, $\sigma^2_{\nu}=\langle\nu^2\rangle - \langle\nu\rangle^2$. \\
Now we estimate the average input to a coding neuron of $\popII$. The neuron receives signals from neurons of $\popI$ coming from both stabilized and baseline connections.
If $\C$ is the number of incoming connections, the average number of high-rate presynaptic neurons will be $\alpha_1 \C$, while those with low rate will be, on average, $\C' = \C (1 - \alpha_1)$.
Since the input pattern used here for testing is identical to the corresponding training pattern, the $\alpha_1 \C$ connections coming from high-rate neurons will certainly be stabilized. The remaining $\C'$ connections come from neurons of $\popI$ at a low rate, however, they may have been stabilized in one of the other $\T - 1$ training patterns. The average number of stabilized connections from low-rate neurons can be calculated using Equation \eqref{eq:av_k}:
\begin{equation}
\label{eq:av_k1}
    \langle k' \rangle = \C' p_{\T -1} = \C(1-\alpha_1) p_{\T -1} ,
\end{equation}
where $p_{\T -1}$ represents the probability shown in Equation \eqref{eq:p} but calculated for $\T -1$ examples. 
For $\alpha_1\alpha_2 \ll 1$, we can observe from Equation \eqref{eq:p} than $p_{\T -1 } \simeq p_{\T}$ and thus
\begin{equation}
\label{eq:av_k1_approx}
\langle k' \rangle \simeq \C(1-\alpha_1) p_{\T} = \langle k \rangle (1 - \alpha_1).
\end{equation}

Formalizing what we just discussed and using the definition of $\langle \rl \rangle$ and $\langle \rh \rangle$ given by Equation \eqref{eq:av_rate_cont} we can write
\begin{equation}
\label{eq:s2_cont}
\begin{split}
    \langle \SII \rangle &= \Wc \alpha_1 \C \langle \rh \rangle + \Wc \langle k' \rangle \langle \rl \rangle + \Wb (\C' - \langle k' \rangle) \langle \rl \rangle=\\
    &= \Wc \alpha_1 \C \langle \rh \rangle + \Wc \langle k \rangle (1 - \alpha_1) \langle \rl \rangle + \Wb (\C - \langle k \rangle)(1 - \alpha_1) \langle \rl \rangle=\\
    &= \Wc \alpha_1 \C \langle \rh \rangle + \Bigl[ (\Wc - \Wb) \langle k \rangle + \C \Wb \Bigr] (1-\alpha_1) \langle \rl \rangle ,
\end{split}
\end{equation}
where we used the expression of $\langle k' \rangle$ and the same approximation $p_{\T-1} \simeq p_{\T}$ shown in Equation \eqref{eq:av_k1_approx}. Indeed, this equation does not take into account the rewiring process, but only the effect of stabilization. Please see Section \ref{subsec:rewiring} for a derivation of $\langle \SII \rangle$ which takes into account both the effects of structural plasticity. We identify this case as "with rewiring" to distinguish it from the case in which non-stabilized connections are not pruned and rewired. Indeed, this distinction is useful to estimate the contribution of this mechanism on $\langle \SII \rangle$.

\subsection{\label{subsec:poiss-indegree} Poisson distribution of incoming connections per neuron}

Hitherto we considered a model in which each neuron of $\popII$ has a fixed number of incoming connections, i.e., a fixed in-degree, $\C$. However, a more general and more realistic approach would consider $\C$ as a variable across the neurons of $\popII$ according to an appropriate probability distribution $P(\C)$.
Here we focus on the case where the number of incoming connections follows a Poisson distribution (i.e. a Poisson-indegree connection rule), although the approach we present here can be easily extended to other distributions.
The values of $\langle \SII \rangle$ and $\langle \Sb \rangle$, previously averaged over the rate $\nu$ and the number of stabilized connections $k$, should be also averaged over the number of incoming connections, so that

\begin{equation}
    \begin{split}
        \langle \langle \Sb \rangle_{\nu,k} \rangle_{\C} &= \sum_{\C} P(\C) \langle \Sb \rangle_{\nu,k}\\
        \langle \langle \SII \rangle_{\nu,k} \rangle_{\C} &= \sum_{\C} P(\C) \langle \SII \rangle _{\nu,k}
    \end{split}
    \label{eq:averaging_over_C}
\end{equation}

where $\langle \Sb \rangle _{\nu,k}$ is given by Equation \eqref{eq:av_sb_cont} and $\langle \SII \rangle_{\nu,k}$ is given by Equation \eqref{eq:s2_cont}. Since these equations are linear in $\C$ and since $\sum_{\C} \C P(\C) = \langle\C\rangle$, Equations \eqref{eq:av_sb_cont} and \eqref{eq:s2_cont} would show $\langle\C\rangle$ instead of $\C$ when averaged over the number of incoming connections per neuron.\\
The variance can be obtained from the equation: 
\begin{equation}
\text{Var}(\langle  \Sb \rangle_{\nu,k, \C})= \sigma^2_{\nu,k,\C} =\langle\Sb^2\rangle_{\nu,k,\C}-(\langle\Sb\rangle_{\nu,k,\C})^2 .
\end{equation}
Knowing that $\langle \varSb\rangle_{\C}=\langle \langle
\Sb^2\rangle_{\nu,k}-\langle\Sb\rangle_{\nu,k}^2\rangle_{\C}
=\langle\Sb^2\rangle_{\nu,k,\C}-\langle\langle\Sb\rangle_{\nu,k}^2\rangle_{\C}$ and that $\langle k \rangle = p\C$ we can write
\begin{equation}
\begin{split}
    \sigma^2_{\nu,k,\C} &= \langle\varSb\rangle_{\C} + \langle\langle\Sb\rangle_{\nu,k}^2\rangle_{\C}-(\langle\Sb\rangle_{\nu,k,\C} )^2=\\
    &= \langle\varSb\rangle_{\C} + \Bigl\{ \langle\nu\rangle \Bigl[ \Wb + p(\Wc-\Wb) \Bigr] \Bigr\}^2 \Bigl[ \langle\C^2\rangle - \langle\C\rangle^2 \Bigr] =\\
    &= \langle\varSb\rangle_{\C} + \langle\nu\rangle^2 \Bigl[ \Wb + p(\Wc-\Wb) \Bigr]^2 \sigma_{\C}^2 .
\end{split}
\label{eq:var_sb_c_variable}
\end{equation}

\subsection{\label{subsec:rewiring} Connection Rewiring}
\begin{figure}[H]
\centering
\includegraphics[width=0.80\columnwidth]{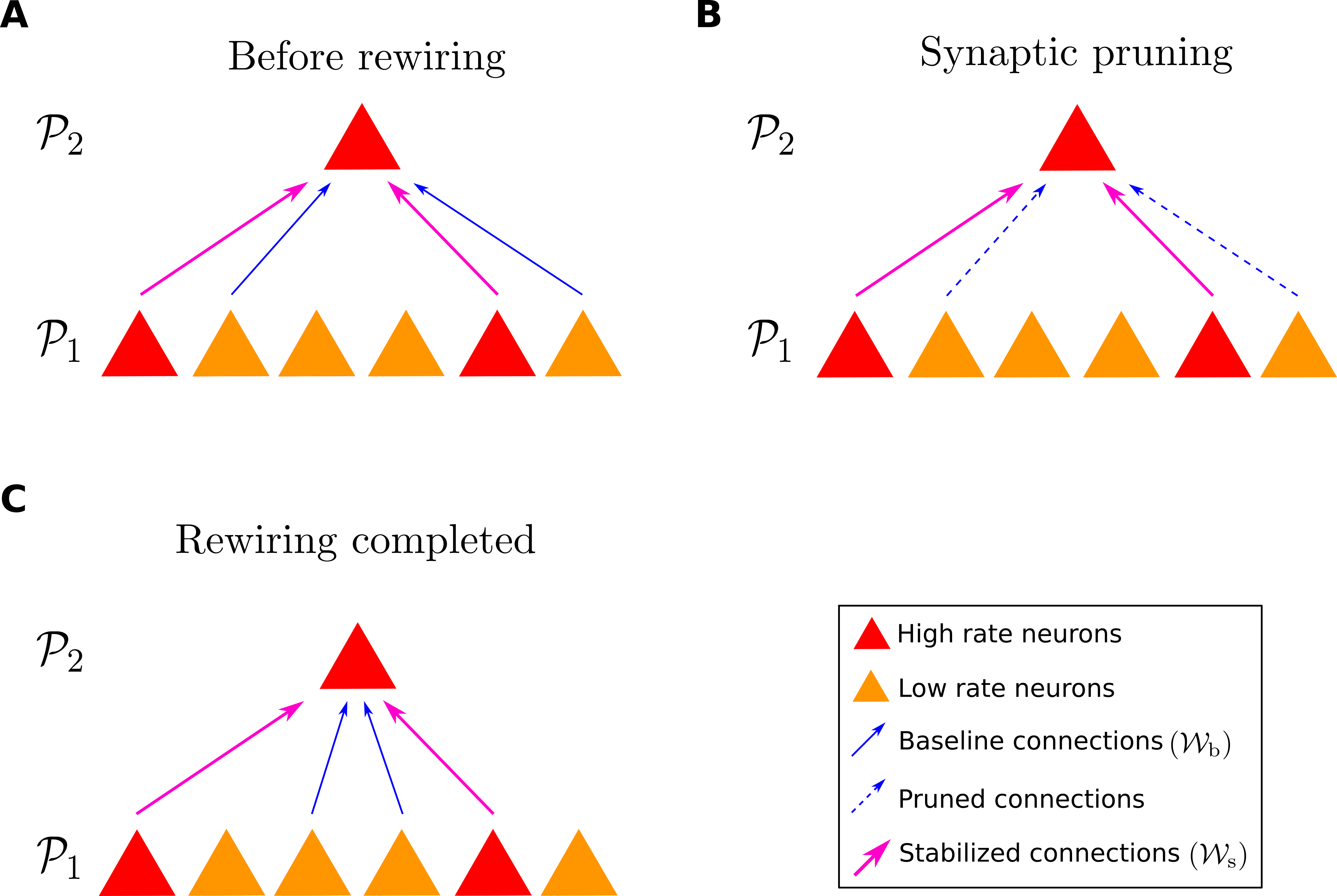}
\caption{\small
     Diagram of the rewiring process. Red and amber triangles represent high and low rate neurons with average rate $\langle \rh \rangle$ and $\langle \rl \rangle$, respectively. Blue and fuchsia connections represent baseline connections with a weight of $\Wb$ and  stabilized connections with a weight of $\Wc$, respectively.
    \textbf{(A)} Diagram representing the connectivity between a subset of $\popI$ neurons and a $\popII$ neuron after synaptic stabilization and before rewiring. \textbf{(B)} Synaptic pruning removes non-stabilized connections every $r$  steps. Pruned connections are depicted as dashed arrows. \textbf{(C)} After pruning, new connections are created, having a randomly chosen presynaptic neuron. 
\label{fig:rewiring}}
\end{figure}

In the proposed approach, rewiring is implemented by periodically pruning unstabilized connections and creating new ones. These procedures are performed with a fixed step on the number of training examples, which we will call {\it rewiring step}, denoted by the letter $r$.
The creation of the new connections is made in such a way as to keep the distribution of the number of incoming connections per neuron unchanged.
If $k$ is the number of stabilized incoming connections of a neuron of $\popII$, after pruning all the non-stabilized connections, $\C - k$ new connections will be created. $\C$ is a fixed number if the fixed-indegree connection rule is used, while it is extracted from a Poisson distribution if the Poisson-indegree rule is selected; in both cases, the presynaptic neurons are randomly extracted from $\popI$. For this reason, rewiring leaves the expressions of the background signal and of the variance on this signal unchanged, while, as we will see, it modifies the input signal to coding neurons.
A diagram of the rewiring process is shown in Figure \ref{fig:rewiring}, which illustrates the activity of a high-rate neuron of $\popII$ and of the presynaptic neurons of its incoming connections in a training example, and the effect of connection rewiring. 

The average number of incoming connections that are stabilized in the current example is equal to the average number of high-rate presynaptic neurons, $\alpha_1 \C$.
The average number of incoming connections that are stabilized in other examples after the entire training,
$\langle k' \rangle$, is given by Equation \eqref{eq:av_k1}:
\begin{equation}
    \langle k' \rangle \simeq p_{\T} \C(1-\alpha_1).
\end{equation}
Let $t$ be the next training index for which rewiring will be applied, and $k'_t$ the number of connections from low-rate neurons that are stabilized before $t$.
These connections will not be affected by rewiring, so even in the test phase with the same input pattern they will have low-rate presynaptic neurons.
The average value of $k'_t$ is
\begin{equation}
\label{eq:avg_k1t}
\langle k'_t \rangle = p_t \C ( 1 - \alpha_1 ),
\end{equation}
where $p_t$ is given by an expression analogous to the one obtained for $p_{\T}$ (Equation \eqref{eq:p})
\begin{equation}
\label{eq:pt}
p_t = 1 - (1 - \alpha_1 \alpha_2)^t .
\end{equation}
On the other hand, there will be $k' - k'_t$ connections displaced by rewiring and stabilized in training examples of index greater than $t$.
Putting all the contributions together, we obtain the following expression for $\SII$:

\begin{equation}
\label{eq:s2_cont_rewiring}
\begin{split}
    \langle \SII \rangle &=
    \alpha_1 \C \Wc \langle \rh \rangle +
    ( \langle k' \rangle - \langle k'_t \rangle )
    \Wc \langle \nu \rangle
    + \langle k'_t \rangle \Wc \langle \rl \rangle
    + \Wb (\C - \alpha_1 \C - \langle k' \rangle)
    \langle \nu \rangle =\\
    &= \alpha_1 \C \Wc \langle \rh \rangle +
    \langle k' \rangle \Wc \langle \nu \rangle +
    \Wb [ \C (1 - \alpha_1) - \langle k' \rangle ]
    \langle \nu \rangle
    - \langle k'_t \rangle \Wc
    ( \langle \nu \rangle - \langle \rl \rangle ).
\end{split}
\end{equation}
To obtain the average value of $\SII$ over all examples, $\langle k'_t \rangle$ must be averaged over all values of the index $t$ for which rewiring is done, i.e.,

\begin{equation}
\label{eq:t_rewiring}
t = r i \qquad i = 0, \dots , \frac{\T}{r} ,
\end{equation}

where $r$ is the rewiring step and for simplicity we assume that $\T$ is a multiple of $r$ and that there is a final rewiring after the last training step. The average of $p_t$ over the rewiring values of $t$ is

\begin{equation}
\langle p_t \rangle = 1 - \frac{b r}{\T + r} ,
\end{equation}

where we introduced a parameter $b$ defined as

\begin{equation}
b = \frac{1 - (1 - \alpha_1 \alpha_2)^{\T + r}}
{1 - (1 - \alpha_1 \alpha_2)^r} .
\end{equation}
The complete derivation is shown in Appendix \ref{app:avg_k1t}.

\subsection{\label{subsec:noise} Introduction of noise into input patterns}
In a realistic learning model, the test patterns will never be exactly the same as the training ones. The ability of a learning model to generalize is linked to the ability to recognize which training pattern or patterns are most similar to a given test pattern, according to appropriate metrics.
To study the generalization capability of the model proposed in this work, the test input patterns were generated starting from the corresponding training input patterns by adding
noise, which is represented by a deviation extracted from a given probability distribution with assigned standard deviation.
In Appendix \ref{app:noise} we describe the effect that noise with a truncated Gaussian distribution has on the firing rates and on the variables $\Sb$, $\SII$, $\varSb$, and SDNR, and we derive the modified equations.

\section{Computational simulations of the model}
The validation of the equations derived in the previous sections was done through simulations with firing-rate-based neuronal network models.
The code of the simulator was written in C++
programming language compiled using the GCC compiler (\url{https://github.com/gcc-mirror/gcc}) (version 10.2.0) and with the GSL (\url{https://www.gnu.org/software/gsl/}) (version 2.7) scientific libraries.
The simulations have been performed using the supercomputers Galileo 100 and JUSUF \citep{VonStVieth2021}.
The networks used for the simulations are generated according to the selected connection rule. In particular, in the case of the fixed-indegree rule, $\C$ incoming connections are created for each neuron of the $\popII$ population, where $\C$ has a fixed value.
In the case of the Poisson-indegree rule, for each neuron of the population $\popII$ the number of incoming connections $C$ is extracted from a Poisson distribution with mean $\langle C \rangle$.
In both cases, the indexes of the presynaptic neurons are randomly extracted on the $\popI$ population. 
The connection weights are initially set to the
baseline value, $\Wb$.
Each training input pattern of the model is generated by extracting, for each neuron of $\popI$, a random number $\nu$ from a lognormal distribution; if $\nu<\rtI$ the rate of the neuron is considered high-rate, otherwise it falls in the low-rate regime.
An analogous procedure is used to generate the corresponding contextual stimulus pattern on the neurons of the population $\popII$ (using a rate threshold $\rtII$).
A connection is stabilized in a training example if both the presynaptic and the postsynaptic neuron are in the high-rate regime.
Connection rewiring is performed every $r$ training steps, as described in Section \ref{subsec:rewiring}.
The test set is generated by randomly extracting
$V$ input patterns from the train set. The patterns of the test set are altered by adding noise extracted from a truncated Gaussian distribution with zero mean.

To estimate $\langle\Sb\rangle$ and $\langle\SII\rangle$ we compute the input of each $\popII$ neuron as the sum of the rate of the presynaptic neurons of its incoming connections multiplied by the synaptic weights (i.e., $\Wc$ or $\Wb$). 
The variance $\varSb$ is evaluated by the formula:
\begin{equation}
    \varSb = \langle\Sb^2\rangle - \langle\Sb\rangle^2 ,
\end{equation}
where the mean values are calculated over the input signals to all the background neurons of $\popII$.
This way, it is possible to obtain the SDNR according to Equation \eqref{eq:SDNR}.

\section{\label{sec:results} Results}
This section compares the results of the simulations of the firing rate model with the theoretical predictions described in Section \ref{sec:rate-cont}. We present the comparison between the theoretical values of the average input signal to background neurons $\langle\Sb\rangle$, the average input signal to coding neurons $\langle\SII\rangle$, the variance of the background input signal $\varSb$ and the signal-difference-to-noise ratio with the values obtained from the simulations.
This way, we are able to assess the capability of the population $\popII$, and thus of the network, to recognize a pattern memorized during the training phase by computing the SDNR using the quantities described above. Lastly, by setting a threshold to the SDNR, we can extract the maximum amount of patterns the network can correctly recall, i.e., the memory capacity.\\
Here, we present simulation results with a Poisson-driven number of incoming connections, with $\langle\C\rangle=5000$. We opted for such an approach since it is more realistic than adopting a fixed amount of connections per neuron. Additionally, the rewiring mechanism is always performed with a rewiring step $r=100$, when not explicitly specified. Each simulation is repeated $10$ times using a different seed for random number generation to ensure the robustness of the simulation results. The values shown in the plots are a result of averaging over the different seeds.

\subsection{Comparison between theoretical predictions and simulation results}
\begin{figure}[h]
\centering\includegraphics[width=\columnwidth, ]{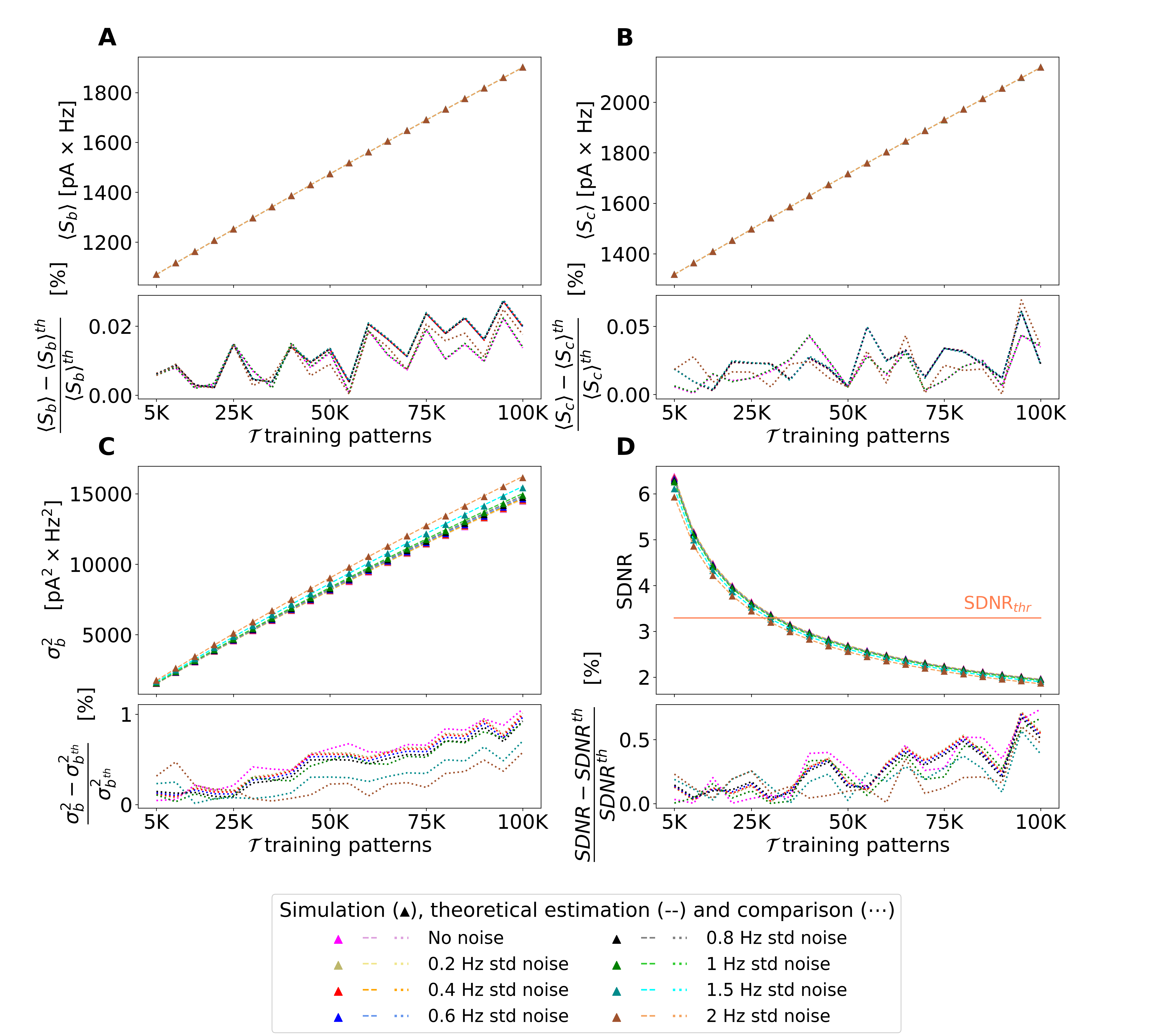}
	\caption{\small 
    Values of $\langle\Sb\rangle$ \textbf{(A)}, $\langle\SII\rangle$ \textbf{(B)}, $\varSb$  \textbf{(C)}, SDNR \textbf{(D)} and percent errors with respect to the theoretical prediction, as a function of the number of training patterns $\T$. 
    Bigger subplots represent the values of the quantities considered as a function of $\T$ for different noise levels, whereas each smaller subplot represents the percentage error of the values shown in the upper subplot. The different color families identify the simulation and theory results when no noise is provided (magenta-pink), or having a noise standard deviation of $0.2$\,Hz (dark khaki-khaki), $0.4$\,Hz (red-orange), $0.6$\,Hz (blue-light blue), $0.8$\,Hz (black-grey), $1$\,Hz (green-light green), $1.5$\,Hz (dark cyan-cyan) and $2$\,Hz (sienna-sand). Because the noise has zero mean, the lines corresponding to the theoretical values of $\langle\Sb\rangle$ and $\langle\SII\rangle$ coincide, with noise having an impact only in the evaluation of the variance of the background signal. The orange horizontal line in panel \textbf{(D)} represents the minimum SDNR for the network to be able to correctly recall the patterns during test. 
	\label{fig:relative-error}}
\end{figure}

To provide a quantitative estimation of the discrepancy between the theoretical predictions and the simulations, we evaluate their relative error, using the theoretical values as a reference.\\
The first study we present is oriented towards the estimation of these parameters as a function of the number of training patterns $\T$. As the number of training patterns increases, so does the number of patterns encoded by each neuron. Since $\alpha_2$ is the probability that a neuron of $\popII$ is in a high-rate level for a single training pattern, on average such neuron will encode $\alpha_2 \T$ patterns of the entire training set. This multiple selectivity of individual neurons is also present in biological neural networks, in which the same neuron can be selective for several stimuli \citep{Rigotti2013}.\\
The test set consists of $V = 1000$ input patterns, generated as described in Section \ref{sec:rate-cont}. Thus, the simulation outcome used for our analysis is an average over the entire test set of the $\Sb$, $\SII$, $\varSb$ and SDNR values obtained for each test pattern. As described previously, the test input patterns are altered from the corresponding training input patterns by adding noise extracted from a truncated Gaussian distribution, with assigned standard deviation. In this section, we present simulation results and comparisons with theoretical predictions for standard deviation values ranging from $0.2$\,Hz to $2$\,Hz. The choice of these values is related to the average of the firing rate distribution of the neuron populations when a pattern is provided to the network. Being the firing rate distributed as a lognormal, we set the highest value of standard deviation to the same value of $\langle\rl\rangle$ (i.e., $2$\,Hz), which is near the average firing rate of the whole neuron population. A larger noise would result in fluctuations significantly larger than the average rate of $\popI$ when the pattern is injected, thus critically altering the rate distribution.\\
Figure \ref{fig:relative-error} shows the curves obtained using different values for the standard deviation of the noise, together with the relative error with respect to theoretical predictions. Moreover, it also shows a set of results in which the test patterns are not altered by noise in order to notice the difference that the noise addition makes in the model results. It can be observed that the curves obtained from the simulations are compatible with the theoretical ones for all the noise levels.

Regarding $\langle\Sb\rangle$ and $\langle\SII\rangle$, the curves corresponding to different noise levels appear perfectly superimposed.
This is due to the fact that the noise is driven by a distribution with zero mean, and thus the addition of noise to the quantities represented in the curves does not alter their average (see Appendix \ref{app:noise} for the details). Regarding $\varSb$, the values corresponding to different noise levels differ from each other and increase with the standard deviation of the noise, in agreement with the theoretical model.\\
The relative error between simulation results and theoretical prediction is quite small: for $\langle\Sb\rangle$ and $\langle \SII \rangle$ the errors span between $0.01$\% and $0.05$\%, whereas $\varSb$ shows a relative error of around $1$\% for all the simulations performed with a different number of training patterns. The orange horizontal line in Figure \ref{fig:relative-error}D represents $\text{SDNR}_{\text{thr}}$, i.e., the minimum value of SDNR that is needed for the network for reliably recalling a pattern, and it is derived using Equation \eqref{eq:sdnr_probability_correct_recall} (see Appendix \ref{app:sdnr_memory_capacity} for additional details). From that, it is possible to derive the network memory capacity $\T_{\text{max}}$. We can notice that depending on the noise level, $\T_{\text{max}}$ spans from around $28000$ when the noise has a standard deviation of $2$\,Hz to more than $30000$ when no noise is applied.

\begin{figure}[h]
\centering\includegraphics[width=\columnwidth]{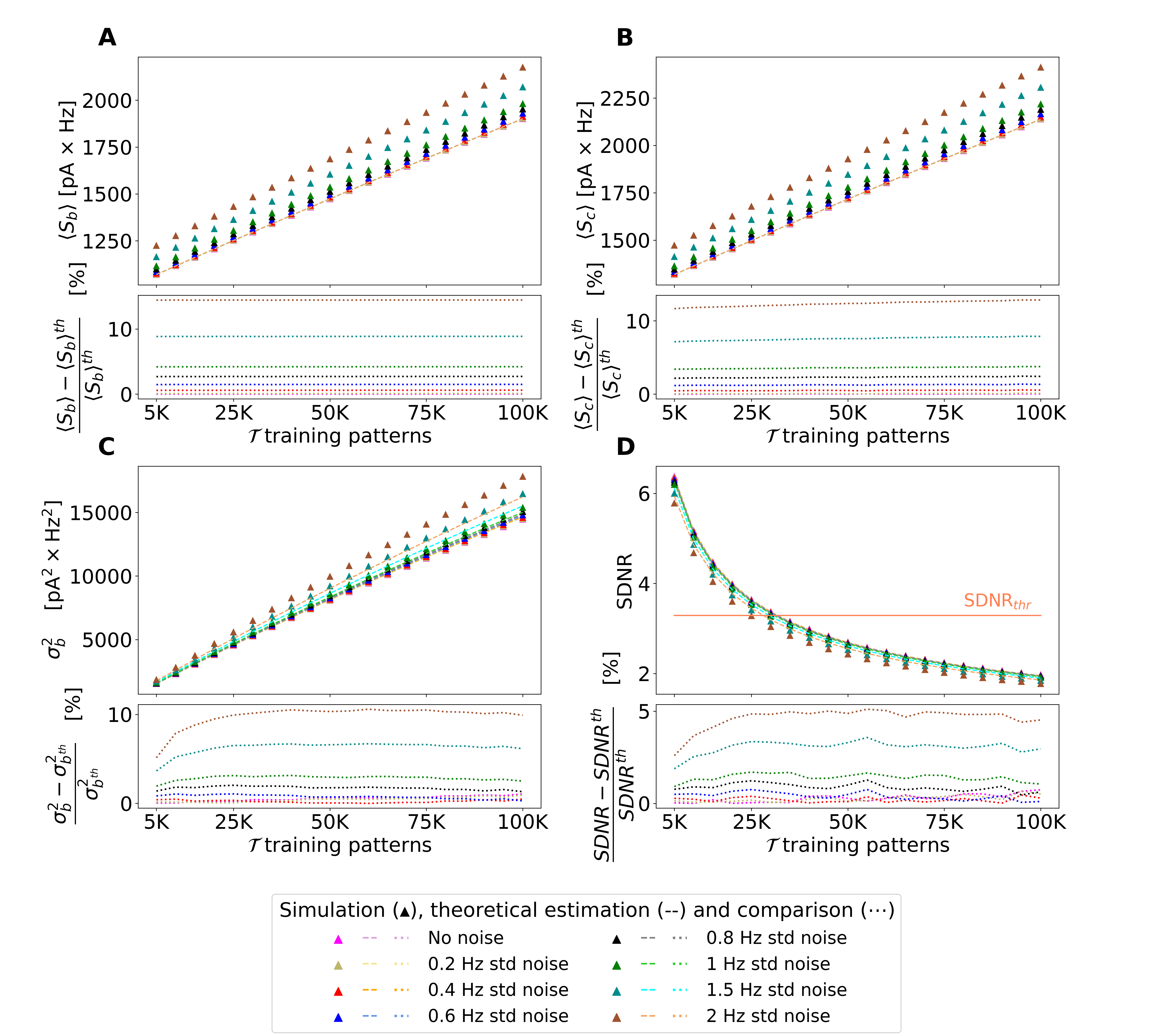}
	\caption{\small 
	Values of $\langle\Sb\rangle$ \textbf{(A)}, $\langle\SII\rangle$ \textbf{(B)}, $\varSb$  \textbf{(C)}, SDNR \textbf{(D)} and percent errors with respect to the theoretical prediction, as a function of the number of training patterns $\T$ when negative rates due to noise addition are saturated to zero.
    The different color families identify the simulation and theory results when no noise is provided (magenta-pink), or having a noise standard deviation of $0.2$\,Hz (dark khaki-khaki), $0.4$\,Hz (red-orange), $0.6$\,Hz (blue-light blue), $0.8$\,Hz (black-grey), $1$\,Hz (green-light green), $1.5$\,Hz (dark cyan-cyan) and $2$\,Hz (sienna-sand). The orange horizontal line in panel \textbf{(D)} represents the minimum SDNR for the network to be able to correctly recall the patterns during test. 
	\label{fig:relative-error-saturation}}
\end{figure}

However, the addition of noise with fluctuations greater than or comparable to the average firing rate can produce negative rate values for a fraction of the neurons. Considering that negative rate values are not physically possible, this behavior can be corrected in the simulations by simply replacing negative values of the firing rates with zero, i.e. saturating negative rates to zero. This correction is equivalent to having a signal-dependent noise distribution, with an average value greater than zero. This negative rate correction has been applied in the simulation, however, the current theoretical model is not able to take this effect into account. Since negative values are replaced by zeros, we would expect the average values of $\Sb$ and $\SII$ evaluated by the simulations that exploit saturation to be greater than the values predicted by the theoretical model.
Figure \ref{fig:relative-error-saturation} shows the behavior of the model with this correction on the neurons firing rate.

As can be seen from the figure, the discrepancies between simulations and theoretical predictions are much higher and can arrive at $10$\%. We also notice that the level of noise with a standard deviation smaller than the average rate does not give rise to significant discrepancies between simulations and theoretical estimations, also because the number of neurons whose firing rate is saturated to zero is smaller. Moreover, it can be seen that a higher noise level yields a smaller SDNR and, consequently, a lower memory capacity. For instance, the memory capacity with $2$\,Hz of noise standard deviation is around $25000$ with saturation enabled, whereas without saturation is around $28000$.

Lastly, in  Figure \ref{fig:relative-error} the relative error of $\varSb$ is greater than that shown for $\langle \Sb \rangle$ and $\langle \SII \rangle$. This is due to a simplification used in theoretical derivation to derive the expression of the variance.
The values of $\Sb$ from which we compute the variance are obtained by incoming connections 
from neurons of $\popI$, but since connections are created randomly, different neurons of the $\popII$ population may have presynaptic neurons in common, and therefore their input signals are correlated. The theoretical model does not take this correlation into account.
However, in Appendix \ref{app:CsuN} we show that the values adopted here make the bias due to this simplification negligible.

\subsection{Impact of synaptic rewiring}
\begin{figure}[h]
    \centering\includegraphics[width=\columnwidth]{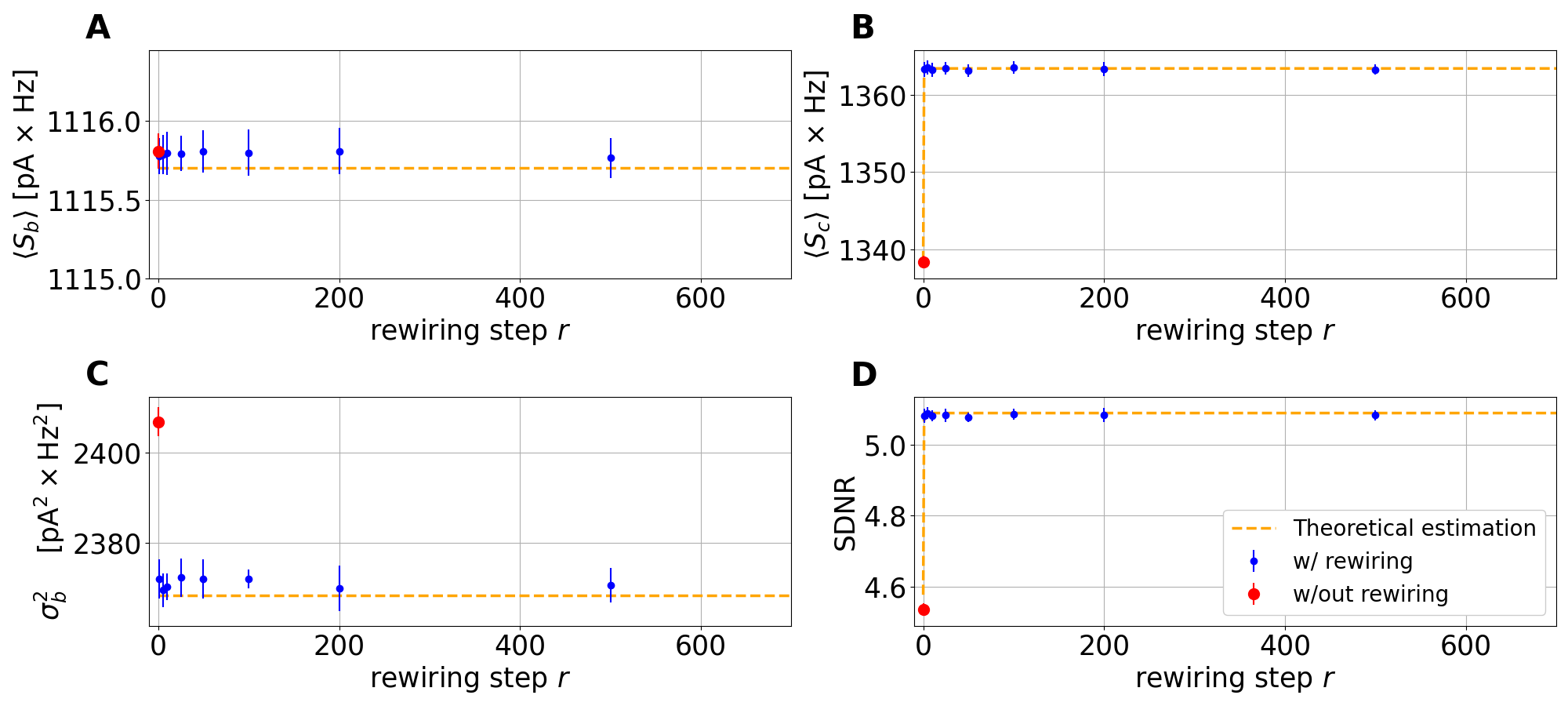}
	\caption{\small 
    Values of $\langle\Sb\rangle$ \textbf{(A)}, $\langle\SII\rangle$ \textbf{(B)}, $\varSb$ \textbf{(C)} and SDNR \textbf{(D)} for a network trained with $10000$ patterns as a function of the rewiring step $r$. The simulations used a continuous rate distribution and a noisy input driven by truncated Gaussian distribution with a standard deviation of $1$\,Hz. Red dots indicate the simulation outcome when connection rewiring is disabled, whereas the blue dots show the simulation results with connection rewiring, using different values of $r$. Error bars represent the standard deviation of the mean obtained from $10$ simulations using different seeds for random number generation, and the orange dotted lines represent the theoretical estimations.
	\label{fig:t_study}}
\end{figure}

In the simulations discussed so far, the rewiring mechanism was always performed with
a rewiring step $r=100$.
This means that every $100$ training patterns, all the unstabilized connections are removed, and new connections are created. This operation represents the effect of homeostatic structural plasticity, which aims at keeping the network balanced by reorganizing connections, while activity-dependent structural plasticity focuses on the stabilization of connections.\\
To motivate the choice of this step for connection rewiring, we show here the results for networks trained for $\T=10000$ patterns with a different rewiring step $r$.
We also show the results of a simulation that does not perform rewiring, in order to highlight the different behavior of a network that combines connection stabilization with periodic rewiring and that of a network that exploits only connection stabilization. Figure \ref{fig:t_study} shows the results obtained by these simulations using different rewiring intervals. 

\begin{figure}[h]
    \centering
	\includegraphics[width=\columnwidth]{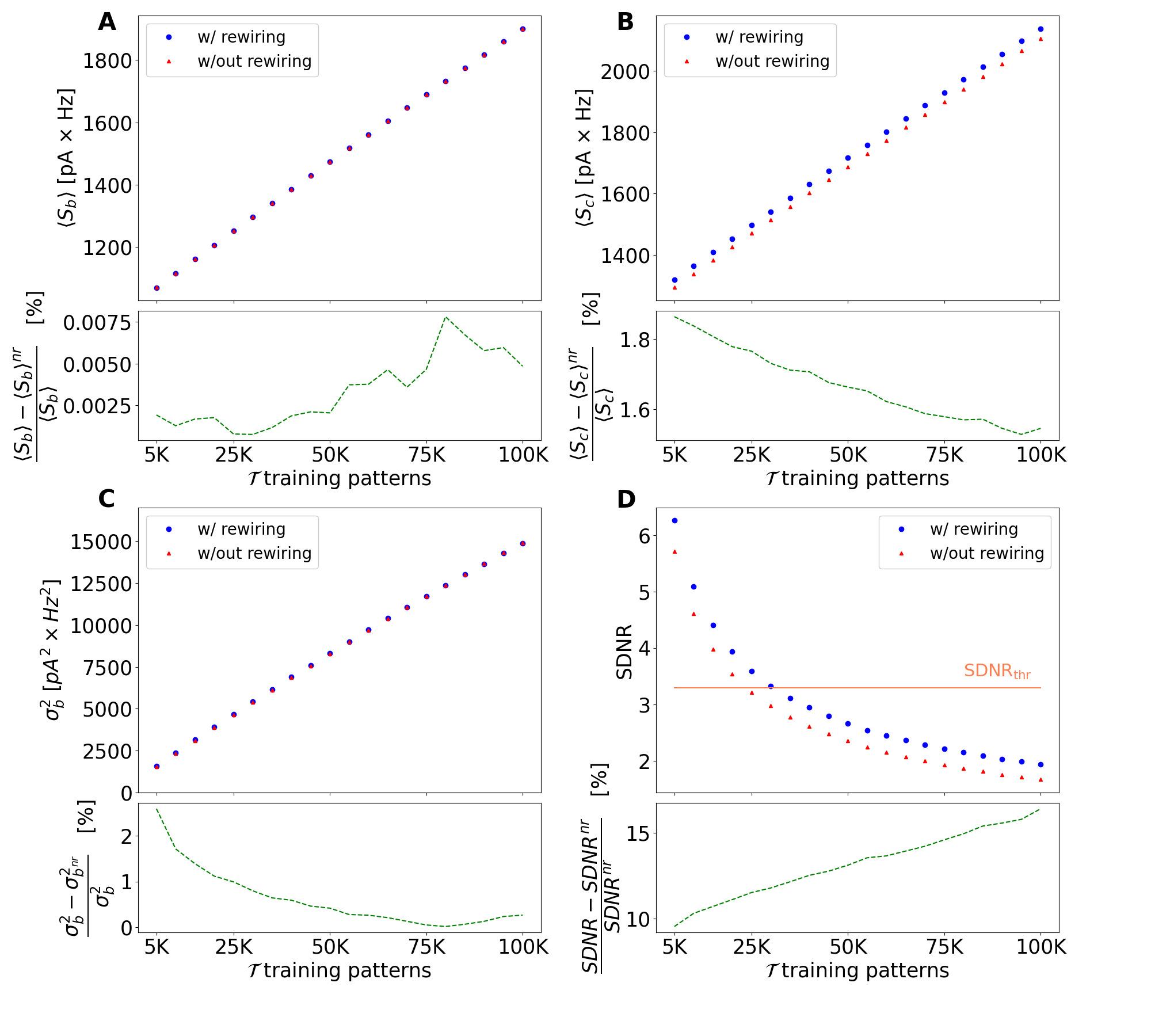}
	\caption{\small 
    Values of $\langle\Sb\rangle$ \textbf{(A)}, $\langle\SII\rangle$ \textbf{(B)}, $\varSb$  \textbf{(C)} and SDNR \textbf{(D)} as a function of the number of training patterns $\T$ when rewiring is performed every $100$ training patterns (blue dots) and when rewiring is disabled (red triangles). Green lines in the sub-panels show the difference between the values obtained with connections rewired or not rewired (indicated with $nr$) as percentage error. The orange horizontal line in panel \textbf{(D)} represents the minimum SDNR for the network to be able to correctly recall the patterns during test. 
	\label{fig:rew_vs_norew}}
\end{figure}

As can be noticed, the values of $\Sb$, $\SII$, $\varSb$ and SDNR do not change significantly as the rewiring step varies.
This means that the value of the step $r$ chosen for the connection rewiring has no substantial impact on the results of the simulations.
On the other hand, some differences emerge when comparing the results of simulations with or without connection rewiring; it can be observed that the signal-difference-to-noise ratio has a lower value when connection rewiring is disabled. This confirms that connection rewiring grants a higher capability of recognizing an input pattern among the several patterns for which the network was trained. Panel C of Figure \ref{fig:t_study} shows also a discrepancy between the simulation without rewiring and the theoretical estimation. Indeed, the theoretical prediction does not show a dependence from the value of $r$, but the simulation results show that when rewiring is not performed, the variance of the background values is higher than expected. This is due to the presence of the bias discussed previously (see also Figure \ref{fig:c_n_comparison} in Appendix \ref{app:CsuN}). Indeed, synaptic rewiring has a similar effect as random repositioning of non-stabilized connections, which reduces such a bias.\\
We also applied a similar protocol for simulations enabling or disabling connection rewiring as a function of the number of training patterns $\T$. The results are shown in Figure \ref{fig:rew_vs_norew}.

We can say that the performance of the model is improved when connection rewiring is enabled, and the relative difference between a rewired or just stabilized connectivity increases when increasing the number of training patterns. Regarding memory capacity, we can notice that the simulation with rewiring reaches the threshold value of SDNR for $\T_{\text{max}}\simeq 30000$ (indeed, it is the same result as the one shown in Figure \ref{fig:relative-error} for $1$\,Hz noise), whereas it decreases to less than $25000$ when rewiring is disabled, resulting in $20\%$ difference in memory capacity.\\
The effect of rewiring can become more relevant when a greater number of connections is stabilized at every step (i.e., with greater values of $\alpha_1$ and $\alpha_2$).
Furthermore, the importance of the rewiring mechanisms can significantly change 
when the average number of connections is not constant, but increases or decreases as a result of rewiring itself. This aspect will be explored in future work.

\section{\label{sec:discussion} Discussion}
In the previous section, the predictions of the theoretical framework have been compared with the results of simulations performed using feed-forward firing-rate-based neuronal networks.
This comparison shows that the proposed framework can accurately predict the values of various relevant quantities for assessing learning and memory capacity in the presence of structural plasticity mechanisms, with differences between theoretical framework and simulation in the order of $1-2$\%.\\
The proposed model offers a detailed framework that incorporates various features observed in biological neural networks. In comparison with the classical Hopfield model, our model considers a more realistic scenario. It accounts for the lognormal distribution of neuron firing rates, random connectivity, connection pruning and rewiring, as well as the potentiation and stabilization of connections between neurons with highly correlated activity.\\
Since the biochemical and biophysical mechanisms underlying structural plasticity are multiple and extremely complex, we opted for a phenomenological approach to capture their main aspects: a simple model of structural plasticity has been exploited, able to represent plasticity processes driven by neuronal activity as well as mechanisms which leads to homeostasis, in agreement with the work of \cite{Fauth2016} which divides structural plasticity mechanisms into these two categories. Structural plasticity driven by neuronal activity is achieved through the potentiation and stabilization of synapses connecting neurons that are concurrently at a high-rate level. This process can be triggered by other forms of plasticity that modify synaptic efficacy, such as STDP, followed by mechanisms involving cytoarchitectural changes, such as the creation of novel connections next to the already existing ones.\\
The homeostatic form of structural plasticity involves a balance between pruning connections that are not utilized over time and the creation of novel connections.
This is achieved in the simulations through periodic connection rewiring, which consists of the removal of non-stabilized connections followed by the creation of new connections. Indeed, potentiation and stabilization not only increase the synaptic efficacy but also prevent connections from being pruned in a subsequent rewiring process. This mechanism can reduce the phenomenon known as catastrophic forgetting, since prevents the deterioration of synaptic contacts already potentiated and stabilized, as also discussed in \cite{Knoblauch2014}. This way, previously learned patterns cannot be forgotten because of further learning and synaptic refinement.\\
This is a simplified approach to model the structural changes that make a connection stable and ensure long-term memory maintenance. Indeed, structural plasticity mechanisms are also present in brain areas that do not have a significant portion of stable synapses. For instance, in regions of the brain such as the hippocampus, long-term memory storage and maintenance occurs despite the absence of a significant fraction of stable connections. This phenomenon is attributed to the remarkable capacity for synaptic turnover and structural plasticity observed in this region \citep{Pfeiffer2018}. Studies have shown that new dendritic spines tend to form close to viable presynaptic terminals, with a higher probability of spine pruning when postsynaptic activity is not correlated with presynaptic activity \citep{Goldin2001}. Therefore, while individual synapses may not persist over time, postsynaptic activity can induce the formation of new dendritic spines near the same site. These newly formed dendritic spines contribute to strengthening the connection between two neurons, facilitating memory retention. In our model, we adopt a simplified but computationally efficient approach to represent these complex dynamics. A single connection can represent multiple synapses between the same pair of neurons, and in the presence of correlated presynaptic and postsynaptic activity, the proliferation of dendritic spines around the same site is depicted by the stabilization and potentiation of this connection.\\
Moreover, we were able to estimate the memory capacity through the evaluation of the SDNR. As shown in Figure \ref{fig:relative-error}, having a feed-forward network of $\NI = \NII = 100000$ neurons, the network can correctly retrieve between $30000$ and $28000$ patterns, depending on the noise level of the patterns provided during test. 

In particular, higher noise provided to the input corresponds to lower values of SDNR, and thus a lower memory capacity. This behavior is also expected from the theoretical model, since noise addition makes the correct pattern retrieval more difficult.  However, fluctuations in the same order, or larger, than the average firing rate can significantly alter the distribution. For this reason, we limited the level of noise to a maximum of $2$\,Hz of standard deviation (i.e., near the average firing rate of the population, which is closer to $\langle\rl\rangle$).

We also studied the behavior of the network in a more realistic condition related to the neurons firing rate as a result of noise addition. Indeed, the noise can lead to neurons showing negative firing rates, which are not biologically possible. Correcting this issue by saturating the negative firing rates to zero, we noticed in Figure \ref{fig:relative-error-saturation} that the SDNR results are slightly lower to the same simulation with rate correction disabled: for instance, when the noise has a standard deviation of $2$\,Hz, the network can correctly retrieve around $25000$ patterns, which is about $12\%$ less than in the case without rate correction. This correction is not implemented in the theoretical framework, and thus we notice higher discrepancies in this case, as also shown in Figure \ref{fig:SDNR_comparison} of Appendix \ref{app:noise}, in which we further increase the noise level to highlight the differences between theory and simulation due to the saturation of negative firing rates. Indeed, a different choice for the values of $\langle \rl \rangle$ and $\langle \rh \rangle$ (and thus a different average rate of the whole distribution) would have an impact on the discrepancies discussed here. In particular, a higher average rate would strongly diminish the amount of neurons having negative firing rate as a result of the noise addition, which saturation is not implemented in the theoretical framework. \\
Whereas Figures \ref{fig:relative-error} and \ref{fig:relative-error-saturation} show the difference in memory capacity due to noise level, Figure \ref{fig:rew_vs_norew} shows the impact of rewiring on the estimation of the memory capacity: this mechanism is able to increase the SDNR between $10\%$ and $15\%$, resulting in the increase of memory capacity by around $20\%$ with respect to the same simulation in which rewiring was disabled. Indeed, synaptic rewiring plays an important role in the refinement of the network, together with the fact that stabilized connections are prevented from being pruned.\\
The framework proposed in this work can be surely extended. It can potentially provide a tool to describe the impact of structural plasticity in cognitive processes such as learning in a large-scale model of the cortex with natural density and plausible characteristics. For instance, the stabilization mechanism can be probability-driven, with a probability depending on the rate of pre- and postsynaptic neurons. This would replace the current deterministic mechanism that requires a firing rate threshold to be exceeded by both neurons to have synaptic stabilization. Moreover, the probability could depend on other variables not necessarily related to the firing rate.
In particular, it has been hypothesized that plasticity mechanisms may also depend on the bursting activity of neurons \citep{Butts2007, Payeur2021}.
The stabilization probability of a connection could therefore depend, in addition to the firing rate of the presynaptic and postsynaptic neurons, also on their bursting activity.\\
While our current model focuses on feed-forward connections between two neuron populations, it does not include self-connections within the target population necessary to describe recurrent network dynamics. Consequently, our model serves as a foundational framework rather than a comprehensive model. Future extensions of our work could involve incorporating self-connections and exploring recurrent network dynamics. Additionally, introducing an inhibitory population could enable the modeling of mechanisms like soft winner-take-all, where competition among neuron groups coding different patterns is mediated by inhibitory signals. Indeed, it is known that the mechanisms of competition through lateral inhibition play a key role in biological learning \citep{Coultrip1992}. These enhancements would enrich our model's capabilities and align it more closely with the complexities of biological neural networks.\\
In this extended model, the theoretical framework can allow to obtain the differential equations governing the dynamics of the activity of the population $\popII$ and the dependence of the coefficients of these equations on the number of training patterns and on the other model parameters \citep{Sergi2023}. Such an extension is currently under development and it will be the subject of future work.\\
Another extension of the model could describe more in detail the mechanisms of synaptic pruning and rewiring. Indeed, connection rewiring as intended in the current model preserves the total number of connections over time, which is a typical behavior of a healthy adult brain \citep{Huttenlocher1979}. However, to shed light on the importance of these mechanisms in neurological disorders, or to perform studies focused on this mechanism in different life stages, this mechanism should be extended to enable different "speed" for the processes embedded in structural plasticity.\\
Moreover, it would be interesting to expand this work through simulations of spiking neural networks, to study learning through structural plasticity in more detail. Indeed, simulators such as NEST \citep{NEST} and its GPU implementation \citep{Tiddia2022} can lead to fast and efficient simulations of large-scale models on supercomputer clusters.\\
In conclusion, this work intends to propose a theoretical framework for learning through structural plasticity. This framework can describe synaptic potentiation, stabilization, pruning and rewiring, and includes several features that can be added in a modular fashion.
The validation has been performed through simulations with firing-rate-based neuronal network models, showing remarkable compatibility between the results of the simulations and theoretical predictions.

\section*{\label{sec:contributions} Author contributions}
\textbf{Conceptualization:} Bruno Golosio, Gianmarco Tiddia and Luca Sergi

\textbf{Data curation:} Bruno Golosio, Gianmarco Tiddia and Luca Sergi

\textbf{Formal analysis:} Bruno Golosio, Gianmarco Tiddia and Luca Sergi

\textbf{Funding acquisition:} Bruno Golosio

\textbf{Investigation:} Bruno Golosio, Gianmarco Tiddia and Luca Sergi

\textbf{Methodology:} Bruno Golosio, Gianmarco Tiddia and Luca Sergi

\textbf{Project administration:} Bruno Golosio

\textbf{Resources:} Bruno Golosio and Gianmarco Tiddia

\textbf{Software:} Bruno Golosio, Gianmarco Tiddia and Luca Sergi

\textbf{Supervision:} Bruno Golosio

\textbf{Validation:} Bruno Golosio, Gianmarco Tiddia and Luca Sergi

\textbf{Visualization:} Bruno Golosio, Gianmarco Tiddia and Luca Sergi

\textbf{Writing - original draft:} Bruno Golosio, Gianmarco Tiddia and Luca Sergi

\textbf{Writing - review and editing:} Bruno Golosio, Gianmarco Tiddia and Luca Sergi

\section*{\label{sec:funding} Funding}
This study was supported by the European Union's Horizon 2020 Framework Programme for Research and Innovation under Specific Grant Agreements No. 945539 (Human Brain Project SGA3), No. 785907 (Human Brain Project SGA2), and by the Italian Ministry of University and Research (MUR) Piano Nazionale di Ripresa e Resilienza (PNRR), project e.INS Ecosystem of Innovation for Next Generation Sardinia – spoke 10 - CUP F53C22000430001 – MUR code: ECS00000038.
We acknowledge the use of Fenix Infrastructure resources, which are partially funded from the European Union's Horizon 2020 research and innovation programme through the ICEI project under the grant agreement No. 800858.\\

\section*{\label{sec:data_av_statement} Data availability statement}
All the simulation code needed to reproduce the results reported in this work, along with the related documentation, is publicly available at Zenodo: \url{https://zenodo.org/doi/10.5281/zenodo.12067622}.

\section*{\label{sec:acknowledgements} Acknowledgements}
We thank Prof. Dr. Paolo Ruggerone and Dr. Ke Zuo for the fruitful discussion on structural plasticity mechanisms.

\nocite{*}
\section{Bibliography}
\bibliography{bibliography}

\newpage
\appendix

\section{Tables and neuron dynamics}
\label{app:tables}

\subsection{Model description}

\begin{table}[H]
\nolinenumbers
\begin{tabular}{
  |@{\hspace*{\marg}}p{0.2\textwidth-2.\marg}@{\hspace*{\marg}}
  |@{\hspace*{\marg}}p{0.8\textwidth-2.\marg}@{\hspace*{\marg}}
  |}
  \hline 
  \multicolumn{2}{|>{\color{black}\columncolor{white}}c|}{\textbf{Summary}}\\
  \hline 
  \textbf{Populations} & $\popI$, $\popII$ \\
  \hline 
  \textbf{Connectivity} & sparse random connectivity\\
  \hline 
  \textbf{Neurons} & firing-rate-based models of point-like neurons
  \\
  \hline 
  \textbf{Synapses} & structural plasticity \\
  \hline 
  \textbf{Input} & firing rate pattern extracted from a probability distribution  \\
  \hline 
\end{tabular}
\begin{tabular}{
  |@{\hspace*{\marg}}p{0.2\textwidth-2.\marg}@{\hspace*{\marg}}
  |@{\hspace*{\marg}}p{0.4\textwidth-2.\marg}@{\hspace*{\marg}}
  |@{\hspace*{\marg}}p{0.4\textwidth-2.\marg}@{\hspace*{\marg}}
  |}
  \hline 
  \multicolumn{3}{|>{\color{black}\columncolor{white}}c|}{\textbf{Populations}}\\
  \hline 
  \textbf{Name} & \textbf{Elements} & \textbf{Size}\\
  \hline 
  $\popI$  & point-like neurons & $\NI$\\
  \hline 
  $\popII$  & point-like neurons & $\NII$\\
  \hline 
\end{tabular}
\begin{tabular}{
  |@{\hspace*{\marg}}p{0.2\textwidth-2.\marg}@{\hspace*{\marg}}
  |@{\hspace*{\marg}}p{0.8\textwidth-2.\marg}@{\hspace*{\marg}}
  |}
  \hline 
  \multicolumn{2}{|>{\color{black}\columncolor{white}}c|}{\textbf{Neuron }}\\
  \hline 
  \textbf{Type} & firing-rate-based neuron model with linear activation function\\
  \hline 
  \textbf{Description} & 
    the state of each neuron is entirely described by the continuous variable $\nu$, which represents its firing rate. We assume that, for a given input pattern to a neuron population, the external stimulus targeting each neuron is stationary.
    Therefore, the neuron firing rate rapidly converges to a steady value. The firing rate of $\popI$ neurons is derived from a continuous distribution.
    During training, the neuron activity of $\popII$ is entirely dependent on the contextual signal (i.e., the contribution of the input signal from $\popI$ is neglected).
    During test, neurons of $\popII$ show a response that depends only on the signal projected by the connections from $\popI$. In this phase, $\popII$ neurons show a linear response (see Section \ref{subsec:neurondynamics} for more details).
    Depending on the firing rate, a neuron can be considered at a low or high rate regime. We define $\rtI$ and $\rtII$ the thresholds that distinguish high and low rate neurons in $\popI$ and $\popII$.
    \\
  \hline 
\end{tabular}
\begin{tabular}{
  |@{\hspace*{\marg}}p{0.1\textwidth-2.\marg}@{\hspace*{\marg}}
  |@{\hspace*{\marg}}p{0.1\textwidth-2.\marg}@{\hspace*{\marg}}
  |@{\hspace*{\marg}}p{0.8\textwidth-2.\marg}@{\hspace*{\marg}}
  |}
  \hline 
  \multicolumn{3}{|>{\color{black}\columncolor{white}}c|}{
  \textbf{Connectivity}
  }\\
  \hline 
  \textbf{Source} & \textbf{Target} & \textbf{Pattern}\\
  \hline
  $\popI$ & $\popII$ & %
                      \begin{itemize}
                      \item the incoming connections are generated by randomly extracting the source neurons from $\popI$; the in-degree (i.e., the number of incoming connections) can be homogeneous, with a fixed number of $\C$ connections per neuron of $\popII$, or driven by a Poisson distribution;
                      \item synaptic weights are $\Wb$ for non-stabilized connections and $\Wc$ for stabilized ones, with $\Wc>\Wb$;
                      \item multiple connections between the same couple of presynaptic and postsynaptic neurons (``multapses'') are allowed by default, but they can be disabled.
                      \end{itemize}\\
  \hline

\end{tabular}

\begin{tabular}{
  |@{\hspace*{\marg}}p{0.2\textwidth-2.\marg}@{\hspace*{\marg}}
  |@{\hspace*{\marg}}p{0.8\textwidth-2.\marg}@{\hspace*{\marg}}
  |}
  \hline 

  \multicolumn{2}{|>{\color{black}\columncolor{white}}c|}{
  \textbf{Synapse}
  }\\
  \hline 
  \textbf{Type} & structural plasticity\\
  \hline 
  \textbf{Description} & initial synaptic weights are set to $\Wb$ for all the instantiated connections;
  when a training pattern is used, considering a connection between a $\popI$ neuron $i$ and a $\popII$ neuron $j$:
 $$\Wb \rightarrow \Wc \text{ if } \nu_\text{i}>\rtI \text{ and } \nu_\text{j}>\rtII ,$$
 i.e., when both presynaptic and postsynaptic neurons are at the high rate regime, the connection is stabilized. Once a connection is stabilized, it cannot return to the initial weight.
  \\
  \hline 
\end{tabular}

\caption{Description of the network model (continued on next page).}
\label{tab:model_description}
\end{table}
\addtocounter{table}{-1}
\begin{table}[H]
\nolinenumbers
\begin{tabular}{
  |@{\hspace*{\marg}}p{0.2\textwidth-2.\marg}@{\hspace*{\marg}}
  |@{\hspace*{\marg}}p{0.8\textwidth-2.\marg}@{\hspace*{\marg}}
  |}
  \hline 

  \multicolumn{2}{|>{\color{black}\columncolor{white}}c|}{
  \textbf{Connection rewiring}
  }\\
  \hline 
  \textbf{Description} & after every $r$ training patterns, non-stabilized connections are pruned, and new connections are created: if $k$ is the number of incoming stabilized connections of a neuron of $\popII$, $\C - k$ new connections will be created, where $\C$ is a fixed number if the fixed-indegree connection rule is used, while it is extracted from a Poisson distribution if the Poisson-indegree rule is selected; in both cases, the presynaptic neurons are randomly extracted from $\popI$.
  \\
  \hline 
\end{tabular}
\begin{tabular}{
  |@{\hspace*{\marg}}p{0.2\textwidth-2.\marg}@{\hspace*{\marg}}
  |@{\hspace*{\marg}}p{0.8\textwidth-2.\marg}@{\hspace*{\marg}}
  |}
  \hline 
  \multicolumn{2}{|>{\color{black}\columncolor{white}}c|}{
  \textbf{Input stimulus}
  }\\
\hline 
\textbf{Description} & firing rate pattern of the neurons of $\popI$ selected from the training or from the test set.\\
\hline 
\end{tabular}
\begin{tabular}{
  |@{\hspace*{\marg}}p{0.2\textwidth-2.\marg}@{\hspace*{\marg}}
  |@{\hspace*{\marg}}p{0.8\textwidth-2.\marg}@{\hspace*{\marg}}
  |}
  \hline 
  \multicolumn{2}{|>{\color{black}\columncolor{white}}c|}{
  \textbf{Contextual stimulus}
  }\\
\hline 
\textbf{Description} & firing rate pattern of the neurons of $\popII$ selected from the training set; used only in the training phase.\\
\hline 
\end{tabular}
\begin{tabular}{
  |@{\hspace*{\marg}}p{0.2\textwidth-2.\marg}@{\hspace*{\marg}}
  |@{\hspace*{\marg}}p{0.8\textwidth-2.\marg}@{\hspace*{\marg}}
  |}
  \hline 
  \multicolumn{2}{|>{\color{black}\columncolor{white}}c|}{
  \textbf{Train set}
  }\\
\hline 
\textbf{Type} & set of $\T$ independent firing-rate patterns of the neurons of $\popI$ (input stimulus) and $\popII$ (contextual stimulus) \\
\hline 
  \textbf{Description} &
  each pattern is randomly generated from predefined firing rate probability distributions (a lognormal probability distribution, in this work). The thresholds $\rtI$ and $\rtII$ are chosen so that the average rate of the neurons belonging to the low or high rate regime corresponds to $\rl$ or $\rh$ respectively, defined as model parameters. The contribution of the signals projected through the connections from $\popI$ to $\popII$ is considered negligible in this case.\\
\hline
\end{tabular}
\begin{tabular}{
  |@{\hspace*{\marg}}p{0.2\textwidth-2.\marg}@{\hspace*{\marg}}
  |@{\hspace*{\marg}}p{0.8\textwidth-2.\marg}@{\hspace*{\marg}}
  |}
  \hline 
  \multicolumn{2}{|>{\color{black}\columncolor{white}}c|}{
  \textbf{Test set}
  }\\
\hline 
\textbf{Type} & set of $V$ firing-rate patterns of the neurons of $\popI$ (input stimulus) \\
\hline 
  \textbf{Description} &
  each pattern is randomly extracted from the train set and altered by a noise, which is modeled by a random deviation extracted from a predefined probability distribution, and added to the firing rate of each neuron. In this work, we model the noise using a truncated Gaussian distribution. The response of the neurons of $\popII$ is entirely dependent on the signals projected through the connections from $\popI$.\\
  \hline
\end{tabular}
\caption{Description of the network model (continued).}
\end{table}

\subsection{Model parameters}

\def\widthA{0.1}
\def\widthB{0.25}
\def\widthC{0.65}
\begin{table}[t]
\nolinenumbers
  \begin{tabular}{
    |@{\hspace*{\marg}}p{\widthA\textwidth-2.\marg}@{\hspace*{\marg}}
    |@{\hspace*{\marg}}p{\widthB\textwidth-2.\marg}@{\hspace*{\marg}}
    |@{\hspace*{\marg}}p{\widthC\textwidth-2.\marg}@{\hspace*{\marg}}
    |}
    \hline 
    \multicolumn{3}{|>{\columncolor{white}}c|}{\textbf{Network and connectivity}}\\
    \hline 
    \textbf{Name} & \textbf{Value } & \textbf{Description}\\
    \hline 
    $\NI$ & $100000$ & number of neurons of $\popI$\\
    \hline 
    $\NII$ & $100000$ & number of neurons of $\popII$\\
    \hline 
    $\C$ & $5000$ & number of connection in-degrees per neuron of $\popII$. In the case of Poisson-driven in-degree, this parameter represents the average of the Poisson distribution.\\
    \hline 
    $\T$ & variable & number of training patterns\\
    \hline 
    $r$ & 100 & number of training patterns between two consecutive connection rewiring\\
    \hline 
    \end{tabular}
  \begin{tabular}{
    |@{\hspace*{\marg}}p{\widthA\textwidth-2.\marg}@{\hspace*{\marg}}
    |@{\hspace*{\marg}}p{\widthB\textwidth-2.\marg}@{\hspace*{\marg}}
    |@{\hspace*{\marg}}p{\widthC\textwidth-2.\marg}@{\hspace*{\marg}}
    |}
    \hline
    \multicolumn{3}{|>{\columncolor{white}}c|}{
    \textbf{Neuron}
    }\\
    \hline 
    \textbf{Name} & \textbf{Value } & \textbf{Description}\\
    \hline 
    $\rl$ & $2.0\sps$ & average firing rate for low rate neurons\\
    \hline 
    $\rh$ & $50\sps$ & average firing rate for high rate neurons\\
    \hline 
    \end{tabular}
  \begin{tabular}{
    |@{\hspace*{\marg}}p{\widthA\textwidth-2.\marg}@{\hspace*{\marg}}
    |@{\hspace*{\marg}}p{\widthB\textwidth-2.\marg}@{\hspace*{\marg}}
    |@{\hspace*{\marg}}p{\widthC\textwidth-2.\marg}@{\hspace*{\marg}}
    |}
    \hline
    \multicolumn{3}{|>{\columncolor{white}}c|}{
    \textbf{Synapse}
    }\\
    \hline 
    \textbf{Name} & \textbf{Value } & \textbf{Description}\\
    \hline
    $\Wb$ & $0.1$\,pA & baseline synaptic weight\\
    \hline
    $\Wc$ & $1$\,pA & stabilized synaptic weight \\
    \hline
    \end{tabular}
  \begin{tabular}{
    |@{\hspace*{\marg}}p{\widthA\textwidth-2.\marg}@{\hspace*{\marg}}
    |@{\hspace*{\marg}}p{\widthB\textwidth-2.\marg}@{\hspace*{\marg}}
    |@{\hspace*{\marg}}p{\widthC\textwidth-2.\marg}@{\hspace*{\marg}}
    |}
    \hline
    \multicolumn{3}{|>{\columncolor{white}}c|}{
    \textbf{Stimulus}
    }\\
    \hline 
    \textbf{Name} & \textbf{Value } & \textbf{Description}\\
    \hline
    $\alpha_1$ & $0.001$ & probability for a neuron of $\popI$ of falling in the high rate regime when an input stimulus is injected\\
    \hline
    $\alpha_2$ & $0.001$ & probability for a neuron of $\popII$ of falling in the high rate regime when a contextual stimulus is injected \\
    \hline
  \end{tabular}
  \caption{Model parameters.}
  \label{tab:model_parameters}
\end{table}

\subsection{Neuron Dynamics}
\label{subsec:neurondynamics}

The dynamics of the neurons evolve according to the following differential equation:
\begin{equation}
\label{eq:firingmodel}
    \tau_i \frac{\nu_i}{dt} = -\nu_i + \Phi \Bigl(\sum_j \mathcal{W}_{ij}\nu_j+ b_i \Bigr)
\end{equation}
where $\nu_i$ is the firing rate of a single neuron $i$, $\tau_i$ is a time constant expressing the time needed for the neuron to reach a steady state firing rate when a constant input is given, $\mathcal{W}_{ij}$ is a matrix containing the synaptic weights, $b_{i}$ is the activation threshold and $\Phi$ represents the activation function. In the asymptotic regime, in which the average firing rates no longer change over time, the previous equation reduces to:
\begin{equation}
    \label{eq:asintotic_condition}
    \nu_i = \Phi \Bigl(\sum_j \mathcal{W}_{ij}\nu_j+ b_i \Bigr)
\end{equation}

To determine the neuron output rate we have to choose an activation function. 

Assuming that the signal is well below saturation, for simplicity, the neuron response can be modeled by a threshold-linear (or ReLU) function
\begin{equation}
\label{eq:ReLu}
\Phi(x) = \mathcal{A} \text{max}\{ 0, x\},
\end{equation}
where $\mathcal{A}$ is a multiplicative coefficient. With this choice,
the average rates of coding and non-coding neurons of $\popII$ can be written as

\begin{equation}
\begin{split}
\langle\nu_\text{c}\rangle &= \langle
    \Phi(\SII + S_\text{o} - S_\text{thresh}) \rangle \\
\langle\nu_\text{nc}\rangle &= \langle
    \Phi(\Sb + S_\text{o} - S_\text{thresh}) \rangle,
\end {split}
\end{equation}

where $S_\text{o}$ is the input signal from (excitatory and/or inhibitory) neuron populations different from $\popI$ and $S_\text{thresh}$ is the activation threshold.
Assuming that the total input signal is above the threshold for both coding and background neurons, the average rates will be linear functions of the input signals:

\begin{equation}
\begin{split}
\langle\nu_\text{c}\rangle &=
    \mathcal{A}(\langle\SII\rangle + \langle S_\text{o}\rangle - S_\text{thresh}) \\
\langle\nu_\text{nc}\rangle &=
    \mathcal{A}(\langle\Sb\rangle + \langle S_\text{o}\rangle - S_\text{thresh}),
\end {split}
\end{equation}

while the variance of the non-coding neuron rate will be

\begin{equation}
\sigma_\text{nc}^2 = \mathcal{A}^2 (\sigma_\text{b}^2 + \sigma_\text{o}^2).
\end{equation}

The SDNR calculated on the rate will therefore be

\begin{equation}
\label{eq:SDNRnu}
    \text{SDNR}_\nu = 
    \dfrac{|\langle\nu_\text{c}\rangle - \langle\nu_\text{nc}\rangle|}
    {\sigma_\text{nc}} =
        \dfrac{|\langle \SII \rangle - \langle \Sb \rangle |}
        {\sqrt{\sigma_\text{b}^2 + \sigma_\text{o}^2}},
\end{equation}

which has an expression similar to that reported in Equation \eqref{eq:SDNR}, with the only difference that there is an additional contribution to the noise due to the signal coming from other populations.

It should also be noted that the definitions of SDNR reported in Equations \eqref{eq:SDNR} and \eqref{eq:SDNRnu} refer to the mean signal difference between single coding and background neurons. Now, we can evaluate the SDNR on the total input signal to coding neurons and an equivalent number of background neurons.
Calling $N_{h,2}$ the mean number of coding neurons in the population $\popII$, we can define

\begin{equation}
\label{eq:SDNRpop}
\text{SDNR}_\text{pop} = 
    \dfrac{|
    N_{h,2}
    \langle \SII \rangle - N_{h,2}
    \langle \Sb \rangle |}
        {\sqrt{N_{h,2}}
        \sigma_\text{b}} =
    \dfrac{\sqrt{\alpha_2 \NII}
    |\langle \SII \rangle -
    \langle \Sb \rangle |}
        {\sigma_\text{b}},
\end{equation}

where we used Equation \eqref{eq:Nh} and $\sqrt{N_{h,2}}\sigma_\text{b}$ is the standard deviation of the total input signal to $N_{h,2}$ non-coding neurons.
Thus, $\text{SDNR}_\text{pop}$ scales with the square root of $\alpha_2 N_2$.

\section{SDNR and memory capacity}\label{app:sdnr_memory_capacity}
In this appendix, we adopt a methodology similar to the one presented in \cite{Schultz:2007}. However, in our model, each neuron's probability of being classified as coding or background is distinct ($\alpha_2$ and $\beta_2$ respectively).  Additionally, the variance of the signal in input on coding neurons may differ from that observed in background neurons. It's also important to note that in \cite{Schultz:2007}, the $\text{SDNR}$ is called the discriminability of the signals and referred to by the symbol $d'$.\\

\begin{figure}[H]
\label{fig:normal_distributions}
\centering
	\includegraphics[scale=0.55]{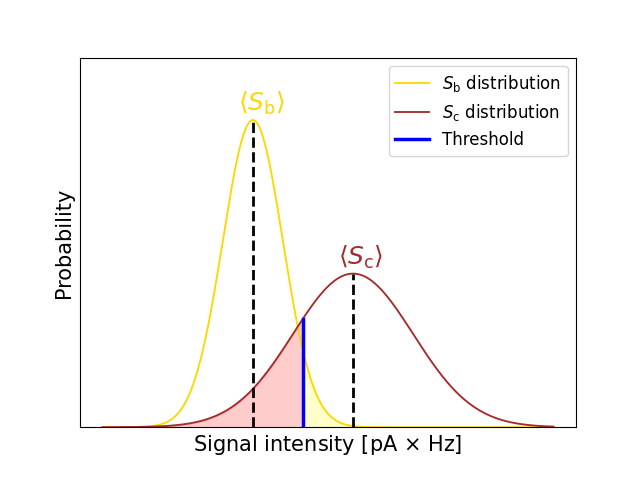}
	\caption{\small 
	Schematic representation of signal distributions for coding neurons (dark red line) and background neurons (gold line). The blue line shows the midpoint of the means of both distributions, used to distinguish coding neurons from background neurons. The shaded red area represents the fraction of neurons mistakenly identified as background, while the yellow area represents the fraction of neurons mistakenly identified as coding. 
	\label{fig:two_gaussian}}
\end{figure}

For simplicity, we assume that the coding and the background neuron signals can be approximated by Gaussian distributions, as shown in Figure \ref{fig:normal_distributions}, and we set a threshold halfway between the averages of the two signals.

By setting a threshold, we can notice in the figure above that a certain fraction of background neurons receive a signal that can be identified as coding. This fraction, identified by the yellow shaded area, represents the false positive rate (FPR) and can be described by the equation
\begin{equation}
 \text{FPR}= \frac{1}{\sqrt{2 \pi \varSb}} \int_{\frac{\langle \SII \rangle + \langle \Sb \rangle}{2}}^{\infty} e^{\frac{(x-\langle \Sb \rangle)^2}{ 2\varSb}}  dx  .
\end{equation}
Reversely, coding neurons that receive a relatively low signal may be identified as background neurons. This fraction, identified by the red shaded area, represents the false negative rate (FNR) and can be described by the equation
\begin{equation}
 \text{FNR}= \frac{1}{\sqrt{2 \pi \varSc}} \int_{-\infty}^{\frac{\langle \SII \rangle + \langle \SII \rangle}{2}} e^{\frac{(x-\langle \SII \rangle)^2}{ 2\varSc}}  dx  . 
\end{equation}

The probability of correct recall $P_\text{C}$ is given by the sum of the rate of correct detections and the rate of correct rejections; equivalently, it is one minus the sum of the rates of false negatives and false positives:
\begin{equation}\label{eq:p_recall}
P_{\text{C}} = 1 - \alpha_2 \text{FNR} - \beta_2 \text{FPR}.
\end{equation}

The prefactors $\alpha_2$ and $\beta_2$ in Equation \eqref{eq:p_recall} account for the different probability of having coding neurons and background neurons. So $P_{\text{C}}$ can be expressed as:

\begin{equation}
\begin{split}
 P_{\text{C}} &= 1 -   \alpha_2 \frac{1}{\sqrt{2 \pi \varSc}} \int_{-\infty}^{\frac{\langle \SII \rangle +\langle \Sb \rangle}{2}} e^\frac{(x-\langle \SII \rangle)^2}{ 2\sigma_{c}^2}dx- \beta_2 \frac{1}{\sqrt{2 \pi \varSb}} \int_{\frac{\langle \SII \rangle + \langle \Sb \rangle}{2}}^{\infty} e^\frac{(x-\langle \Sb \rangle)^2}{ 2 \varSb} dx=\\
&= 1  -\alpha_2 \frac{1}{\sqrt{2 \pi \varSc}} \int_{-\infty}^{\langle \SII \rangle} e^{\frac{(x-\langle \SII \rangle )^2}{2\varSc}}dx+ \alpha_2\frac{1}{\sqrt{2 \pi \varSc}} \int_{\frac{\langle \SII \rangle +\langle \Sb \rangle}{2}  }^{\langle \SII \rangle}  e^{\frac{(x-\langle \SII \rangle)^2}{2\varSc}} dx +\\
& - \beta_2 \frac{1}{\sqrt{2 \pi \varSb}} \int_{\langle \Sb \rangle}^{\infty} e^{\frac{(x-\langle \Sb \rangle )^2}{2\varSb}}dx+ \beta_2\frac{1}{\sqrt {2 \pi \varSb}} \int_{\langle \Sb \rangle}^{\frac{\langle \SII \rangle +\langle \Sb \rangle}{2}}  e^{\frac{(x-\langle \Sb \rangle)^2}{2\varSb}} dx .
\end{split}
\end{equation}
So, using the definition of the error function erf, we obtain:
\begin{equation}
\label{eq:p_recall2}
\begin{split}
    P_\text{C}=& 1-  \frac{1}{2} (\alpha_2 +  \beta_2) + \frac{1}{2}\Bigl[\alpha_2 \text{erf}\Bigl(\frac{\text{SDNR}}{\sqrt{8}}\phi\Bigr) +\beta_2 \text{erf}\Bigl(\frac{\text{SDNR}}{\sqrt{8}}\Bigr)\Bigr]=\\
    =&\frac{1}{2} + \frac{1}{2}\Bigl[\alpha_2 \text{erf}\Bigl(\frac{\text{SDNR}}{\sqrt{8}}\phi\Bigr) +\beta_2 \text{erf}\Bigl(\frac{\text{SDNR}}{\sqrt{8}}\Bigr)\Bigr]. \\
\end{split}
\end{equation}
Where $\phi$ is defined as the ratio between the standard deviation on the input signal of background neurons and the standard deviation on the input signal of coding neurons:
\begin{equation}
    \phi= \frac{\sigma_\text{b}}{\sigma_\text{c}}.
\end{equation}
Simulations show that this value is approximately $\phi \simeq \frac{1}{3}$ (data not shown). The Equation  \eqref{eq:p_recall2} show also that the probability of correct recall depends on the number of neurons through $\alpha_2$ and $\beta_2$, which take into account the size of the populations $\popII$. In the Equation \eqref{eq:p_recall2}, the term with $\alpha_2=10^{-3}$ is negligible, leaving only the term with $\beta_2=(1-10^{-3}) \simeq 1$:

\begin{equation}
    P_\text{C} \simeq  \dfrac{1}{2}\Bigr[1 + \text{erf}\Bigr(\frac{\text{SDNR}}{\sqrt{8}}\Bigr)\Bigr].
\end{equation}

In the graph below we show a plot of the probability of correct recall as a function of the $\text{SDNR}$ using the previous formula.
\begin{figure}[H]
\centering
	\includegraphics[scale=0.55]{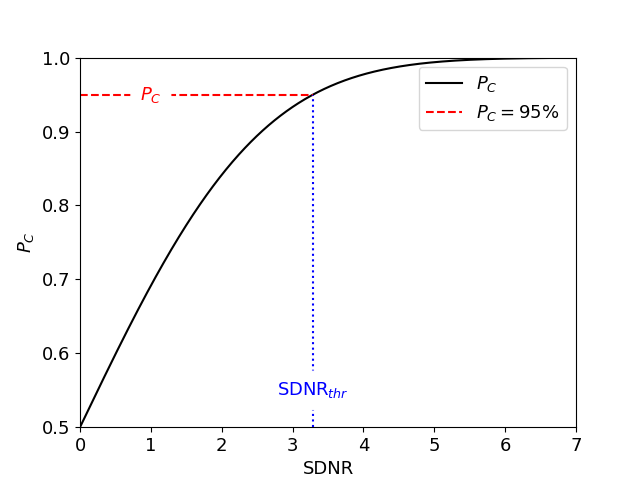}
	\caption{\small 
	Probability of correct recall as a function of the signal-difference-to-noise ratio on the single neuron. The blue dotted line represents the threshold $\text{SDNR}_{\text{thr}}\simeq 3,3 $ corresponding to a $95 \%$ probability (red dotted line). 
 \label{fig:p_recall}}
\end{figure}

By setting a recall probability of $95\%$, as shown in Figure \ref{fig:p_recall}, we obtain a threshold equal to ${\text{SDNR}_{\text{thr}}\simeq 3.3}$. This threshold allows us to calculate the maximum number of patterns stored by our network, as shown in Section \ref{sec:results}, thereby quantifying the memory capacity of our network.

\section{Lognormal distribution of the firing rate}
\label{app:distr_fr}
\begin{figure}[H]
    \centering
    \includegraphics[scale=0.6]{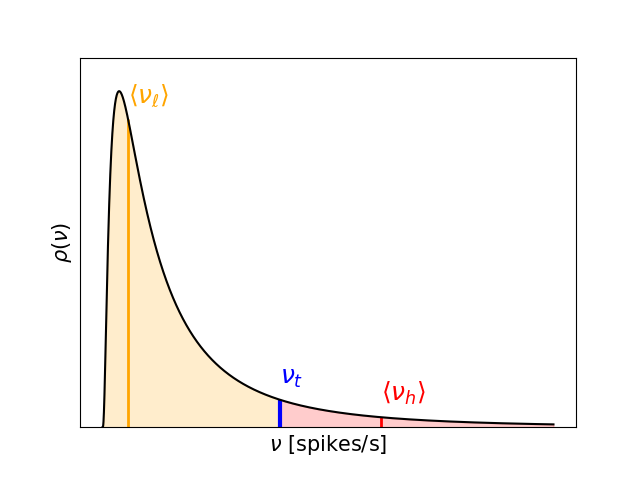}
    \caption{Lognormal distribution of firing rate. The black solid line indicates the probability distribution, which is divided into two sections by the rate threshold $\rt$ (blue, vertical line). The amber band represents the distribution of rate below the threshold, whose mean is $\langle \rl \rangle$ (amber, vertical line). The red band represents the distribution of neurons whose rate is above the threshold. Here the average of this section is $\langle \rh \rangle$ (red, vertical line).}
    \label{fig:lognormal}
\end{figure}

The theoretical framework proposed in this work is valid for a generic firing rate probability distribution. However, the model validation presented in the result section is focused on a lognormal distribution, which is a continuous probability distribution of a random variable $\nu$ whose logarithm 
$\ln(\nu)$ is normally distributed.
The probability density function of this distribution is
\begin{equation}
\label{eq:lognormal_app}
 \rho_\text{LN}(\nu) = \frac{1}{\sqrt{2\pi} \sigma \nu} \cdot \exp\Bigl( -\frac{(\ln(\nu)-\mu)^2}{2\sigma^2}\Bigr),
\end{equation}
where $\mu$ and $\sigma$ are the mean and standard deviation of $\ln(\nu)$.
Expanding Equation \eqref{eq:av_rate_cont} using Equation \eqref{eq:lognormal_app} we have

\begin{equation}
\begin{split}
    \langle \rl \rangle &=\frac{1}{\beta_1}\int_{-\infty}^{y_t} \nu(y)G_{\sigma, \mu}(y) dy=\frac{1}{\beta_1}\int_{-\infty}^{y_t} e^y \frac{1}{\sqrt{2\pi \sigma^2}}e^{\frac{-(y-\mu)^2}{2\sigma^2} } dy\\
    \langle \rh \rangle &= \frac{1}{\alpha_1}\int_{y_{t}}^\infty \nu(y) G_{\sigma, \mu}(y) dy= \frac{1}{\alpha_1}\int_{y_{t}}^\infty e^y \frac{1}{\sqrt{2\pi \sigma^2}} e^{\frac{-(y-\mu)^2}{2\sigma^2} } dy ,
\end{split}
\label{eq:av_rate_cont2}
\end{equation}

where $y$ is a variable representing the logarithm of the firing rate, $y=\ln(\nu)$, and follows a normal distribution $G_{\sigma, \mu}(y)$, while $y_t$ represents the value linked to the threshold value on the rate $\nu_t$ ($y_t=\ln(\nu_t)$). Figure \ref{fig:lognormal} depicts an example of firing rate distribution, with the threshold and the average values of low and high firing rates.

In the logarithmic representation the area of the portion of the Gaussian $G_{\sigma, \mu}(y)$ having $y<y_t$ corresponds to the probability that a neuron has a low rate, $\beta_1 $. Therefore we can write:

\begin{equation}
    \beta_1= \int_{-\infty}^{y_t} G_{\sigma, \mu}(y) dy= \frac{1}{2}+ \int_{\mu}^{y_t} \frac{1}{\sqrt{2 \pi \sigma^2}} e^{- \frac{(y-\mu)^2}{2 \sigma^2}} dy .
\end{equation}

Substituting $x= \frac{y-\mu}{\sqrt{2} \sigma}$ we obtain:

\begin{equation}
    \beta_1=\frac{1}{2}+ \int_{0}^{\frac{y_t-\mu}{\sqrt{2} \sigma}} \frac{1}{\sqrt{\pi}} e^{-x^2} dx = \frac{1}{2}+\frac{1}{2} \text{erf}\Bigl(\frac{y_t-\mu}{\sqrt{2}\sigma}\Bigr),
\end{equation}

where with $\text{erf}(x)$ we indicate the error function, defined as:

\begin{equation}
    \text{erf}(x)= \frac{2}{\sqrt{\pi}} \int_{0}^x e^{-t^2} dt ,
\end{equation}

and then:

\begin{equation}
 y_t=\mu+\sqrt{2}\sigma \text{erf}^{-1}(2\beta_1-1) ,
\end{equation}

where $\text{erf}^{-1}$ is the inverse of the $\text{erf}$ function. By substituting $z=y-\mu$ we can rewrite $\nu_h$ from Equation \eqref{eq:av_rate_cont2} as:

\begin{equation}
\begin{split}
 \langle \nu_h \rangle=&  
  \frac{1}{\alpha_1}\int_{y_{t}-\mu}^\infty  \frac{1}{\sqrt{2\pi \sigma^2}} e^{z+\mu-\frac{z^2}{2\sigma^2} } dz =   \frac{1}{\alpha_1} \frac{e^\mu}{\sqrt{2\pi \sigma^2}}  \int_{y_{t}-\mu}^\infty e^{-\frac{z^2 -2\sigma^2z}{2\sigma^2} } dz= \\
 =&\frac{1}{\alpha_1} \frac{e^\mu}{\sqrt{2\pi \sigma^2}}  \int_{y_{t}-\mu}^\infty e^{-\frac{(z -\sigma^2)^2-\sigma^4}{2\sigma^2} } dz =
     \frac{1}{\sqrt{2\pi}}\frac{e^{\mu+\frac{\sigma^2}{2}}}{\sigma \alpha_1} \int_{y_{t} -\mu}^\infty e^{\frac{-(z-\sigma^2)^2}{2 \sigma^2}}dz .
    \end{split}
\end{equation}

Making a further substitution $\xi=\frac{z-\sigma^2}{ \sqrt{2} \sigma}$ finally we find:

\begin{equation}
\begin{split}
    \langle \nu_h \rangle&= \frac{1}{\sqrt{\pi}} \frac{e^{\mu + \frac{1}{2}\sigma^2}}{\alpha_1} \int_{\frac{y_{t} -\mu-\sigma^2}{\sqrt{2}\sigma}}^\infty e^{-\xi^2}d\xi=\\
     &= \frac{1}{\sqrt{\pi}} \frac{e^{\mu + \frac{1}{2}\sigma^2}}{\alpha_1} \Bigl(\int_{0}^\infty e^{-\xi^2}d\xi- \int_{0}^{\frac{y_{t} -\mu-\sigma^2}{\sqrt{2}\sigma}} e^{-\xi^2}d\xi \Bigr)=\\
     &=\frac{e^{\mu + \frac{1}{2}\sigma^2}}{\alpha_1}\Bigl[\frac{1}{2}- \frac{1}{2} \text{erf} \Bigl(\frac{y_{t} -\mu-\sigma^2}{\sqrt{2}\sigma} \Bigr) \Bigr]=\\
     &=\frac{\langle \nu \rangle }{2 \alpha_1}\Bigl[1- \text{erf} \Bigl(\text{erf}^{-1}(2\beta_1-1) -\frac{\sigma}{\sqrt{2}}\Bigr) \Bigr] ,
    \end{split}
\end{equation}

where with $\langle \nu \rangle$ 
we indicate the average rate, which for the lognormal distribution is given by the known expression

\begin{equation}
\label{eq:avg_nu_lognormal}
\langle \nu \rangle = e^{\mu + \frac{1}{2}\sigma^2} .
\end{equation}

With similar steps we obtain the expression of
$\langle \nu_l \rangle$:

\begin{equation}
    \langle \nu_l \rangle= \frac{\langle \nu \rangle }
    {2 \beta_1}\Bigl[1- \text{erf} \Bigl(\text{erf}^{-1}(2\alpha_1-1) -\frac{\sigma}{\sqrt{2}}\Bigr) \Bigr] .
\end{equation}

From these two equations we can finally derive the relationships between $\sigma$, $\beta_1$, $\nu$ and $\nu_h$ or $\nu_l$ respectively:

\begin{equation}
    \sigma= \sqrt{2} \Bigl[ \text{erf}^{-1} ( 2 \beta_1 - 1) -
    \text{erf}^{-1} \Bigl( 1 -
    {\frac{2 \alpha_1 \langle \nu_h \rangle}
    {\langle \nu \rangle}} \Bigr) \Bigr] ,
\end{equation}

\begin{equation}
    \sigma= \sqrt{2} \Bigl[ \text{erf}^{-1} ( 2 \alpha_1 - 1) -
    \text{erf}^{-1} \Bigl( 1 -
    {\frac{2 \beta_1 \langle \nu_l \rangle}
    {\langle \nu \rangle}} \Bigr) \Bigr] .
\end{equation}

Using the Equation \eqref{eq:avg_nu_lognormal} we can rewrite $\mu$ as:

\begin{equation}
    \mu= \ln(\langle \nu \rangle) -\frac{\sigma^2}{2} .
\end{equation}

The average rate $\langle \nu \rangle$
can also be expressed as a function of
$\langle \rh \rangle$ and $\langle \rl \rangle$:

\begin{equation}
    \langle \nu \rangle = \alpha_1 \langle \rh \rangle +
    \beta_1 \langle \rl \rangle .
\end{equation}

The latter equations allow us to express the parameters of the lognormal distribution $\sigma$ and $\mu$ as a function of the parameters of the model,
$\alpha_1$,  $\langle \rh \rangle$ and $\langle \rl \rangle$.

\section{Estimation of the variance of $k$}
\label{app:var_k}
In this appendix, we will compute the variance on the number of stabilized connections in input to a neuron of $\popII$ (i.e., $\sigma^2_{k}$) which, as we have seen previously, enters the formula for the variance on the background signal. For the calculation, we will use the table below which represents the two states, high rate (1) or low rate (0), for a single neuron of the population $\popII$ and the presynaptic neurons of its input connections in a complete simulation over $\T$ patterns.

\begin{table}[H]
\nolinenumbers
\centering
\resizebox{8cm}{3cm}{%
\begin{tabular}{|| c | c | c | c| c | c | c | c | c ||} 
 \hline
 t & $\mathcal{O}$ & $\mathcal{I}_0$ & $\mathcal{I}_1$ & .... & $\mathcal{I}_{k-1}$ & $\mathcal{I}_{k}$ & .... & $\mathcal{I}_{\C-1}$\\ [0.5ex] 
\hline\hline 
 0 & 1& $x_{00}$& & &  $x_{k-1,0}$& 0& & 0   \\
\hline 
 1 & 1&.. & & & ..&0 & & 0   \\ 
 \hline
 2 & 1&.. & & & ..&0 & & 0 \\
\hline
 .. & .. & ..& & & ..& .. & & ..\\
\hline
 m-1 & 1& $x_{0m}$ & & &  $x_{k-1,m-1}$&0 & & 0\\

\hline
.. &0 & & & & & & & \\
\hline
.. & ..& & & & & & & \\

 \hline
$\T-1$ &0 & & & & & & & \\ [1ex] 
\hline
\end{tabular}}
\vspace{2mm}
    \caption{Table representing
    the two states high rate (1) or low rate (0) for a single neuron of the population $\popII$ and for the presynaptic neurons of its input connections in a complete simulation.    
    Each row represents a training pattern, with index ranging from 0 to $\T-1$. 
    The first two columns represent the training pattern index $t$ and the rate level $\mathcal{O}$, high or low, of the $\popII$ neuron. 
    The other columns $\mathcal{I}_j$ represent the rate level, high or low, of the presynaptic neurons connected to the neuron of $\popII$ through its $C$ incoming connections. The entries for rate levels can be 0 or 1 for low rate and for high rate respectively; in case of continuous distribution of the rate, the two levels correspond to a rate over or under the threshold $\nu_{\text{t}}$. 
    The table shows the case in which the $\PII$ neuron is in the high-rate level for the first $m$ examples and in the low-rate level for $\T -m$ examples, while the last $\C - k$ presynaptic neurons
    are in the low-rate level for the first $m$ examples.
}
    \label{fig:rate_matrix}
\end{table}

Given the scheme of Table \ref{fig:rate_matrix}, we call:
\begin{itemize}
    \item $\alpha_1$: probability that a neuron of $\popI$ is in the high-rate level, i.e. probability that a cell of a column $\mathcal{I}_j$ is equal to one;
    \item $\alpha_2$: probability that the neuron of $\popII$ is in the high-rate level for a given example, i.e., probability that a cell of the column $\mathcal{O}$ is equal to one;
    \item $\alpha_2^m$: probability that the neuron of $\popII$ is in the high-rate level for the first $m$ patterns;
    \item $(1-\alpha_2)^{\T - m}$: probability that the neuron of $\popII$ is in the low-rate level for the remaining $\T - m$ patterns;
    \item $(1-\alpha_1)^m$: probability that a neuron of $\popI$ is in the low-rate level for the first $m$ patterns;
    \item $1-(1-\alpha_1)^m$: probability that a neuron of $\popI$ is in the high-rate level for at least one pattern out of the first $m$;
    \item $[1-(1-\alpha_1)^m]^{k}$: probability that every neuron of $\popI$ of the columns
    $\mathcal{I}_0$, ...., $\mathcal{I}_{k-1}$
    is above threshold for at least one pattern among the first $m$;
    \item $(1-\alpha_1)^{m(\C-k)}$: probability that every neuron of $\popI$ of the last $\C - k$ columns is below threshold for the first $m$ patterns.
\end{itemize}

Now we can combine all these results to calculate the probability that one neuron of $\popII$ and $k$ presynaptic neurons of its input connections are at the high level for $m$ generic patterns (i.e., not necessarily the first $m$). To do this we have to take into account that the neuron of $\popII$ will not necessarily be at the high level in the first $m$ examples and that the neurons of $\popI$ at the high level will not necessarily be the first $k$ (as in the case shown in the table). 
For this, we have to use binomial coefficients that will take into account all possible combinations
in the choice of $m$ patterns out of all possible $\T$ patterns and in the choice of $k$ presynaptic neurons out of a total of $\C$ connections:

\begin{equation}
\label{eq:Q_mk}
    Q(m,k) = \binom{\T}{m} p_{2}^m (1-\alpha_2)^{\T-m}
    \binom{\C}{k}  \hspace{2mm} [1-(1-\alpha_1)^m]^k  \hspace{2mm} (1-\alpha_1)^{m(\C-k)} .
\end{equation}

The probability that $k$ connections of a generic neuron of $\popII$ are stabilized can be calculated by adding $Q(m,k)$ over all possible values of $m$:

\begin{equation}
    P(k) = \sum_{m=0}^\T Q(m,k) ,
\end{equation}

and the average number of stabilized connections can be calculated as:

\begin{equation}
\begin{split}
    \langle k \rangle &= \sum_{m,k}  k Q(m,k) = \sum_m \binom{\T}{m} p_{2}^m (1-\alpha_2)^{\T-m} \sum_{k=0}^{\C} k \binom{\C}{k} (1-\beta_1^m)^k \beta_1^{m (\C-k)} =\\
    &= \sum_m \binom{\T}{m} p_{2}^m (1-\alpha_2)^{\T-m} \C(1-\beta_1^m)
    =\\
    &= \C \Bigl[\hspace{0.2cm}\sum_m \binom{\T}{m} p_{2}^m (1-\alpha_2)^{\T-m}
    - \sum_m \binom{\T}{m} (\alpha_2 \beta_1)^m (1-\alpha_2)^{\T-m} \Bigr]
    =\\
    &= \C \Bigl[\hspace{0.2cm}
    1 - \sum_m \binom{\T}{m} (\alpha_2 - \alpha_1 \alpha_2)^m (1-\alpha_2)^{\T-m} \Bigr] ,
    \end{split}
\end{equation}

where $\beta_1 = 1 - \alpha_1$ and we have used the formula for the mean value of a binomial distribution:

\begin{equation}
\sum_{k=0}^{n} k \binom{n}{k} p^k (1-p)^{n-k}= np .
\end{equation}

Using the relationship:

\begin{equation}
    (a+b)^n=\sum_{k=0}^n \binom{n}{k} a^k b^{n-k} ,
\end{equation}

we can get the expression of $\langle k \rangle$:

\begin{equation}
\label{eq:avg_k}
    \langle k \rangle = \C [1-(1-\alpha_1 \alpha_2)^\T] .
\end{equation}

To calculate $\sigma_{k}^2$ we must also calculate $\langle k^2 \rangle$:

\begin{equation}
\begin{split}
    \langle k^2 \rangle= \sum_{m,n} Q(m,k) k^2 \hspace{2.5cm}\\
    \langle k^2 \rangle= \sum_m \binom{\T}{m} p_{2}^m (1-\alpha_2)^{\T-m} \cdot \sum_{k=0}^C \binom{\C}{k} (1-\beta_1^m)^k \beta_1^{m (C-k)} k^2 .
    \end{split}
\end{equation}

After some calculations, analogous to the case of
$\langle k \rangle $, we obtain the following formula:

\begin{equation}
\label{eq:avg_k2}
    \langle k^2 \rangle= C(C-1) [1+ \alpha_1 \alpha_2 (\alpha_1 - 2)]^{\T}
    - C (2C - 1) (1 - \alpha_1 \alpha_2)^{\T} + C^2 .
\end{equation}

Finally, the variance can be calculated from equations
\eqref{eq:avg_k} and \eqref{eq:avg_k2} as

\begin{equation}
    \sigma_{k}^{2} = \langle k^2 \rangle 
    - \langle k \rangle^2 .
\end{equation}

The values of $\langle k \rangle$ and $\langle k^2 \rangle$ derived in the above formulas represent the expected values of $k$ and $k^2$ over all neurons of $\popII$ for a random input pattern.
The average number of stabilized incoming connections of a background neuron in a given example can be computed by observing that, by definition, such neuron cannot have stabilized connections for the considered example, while the average number of connections stabilized in different examples can be computed by replacing $\T$ with $\T - 1$ in Equation \eqref{eq:avg_k}. Same applies to $\langle k^2 \rangle$. In this work, we assume that $\alpha_1\alpha_2 \ll 1$,
and thus $(1-\alpha_1\alpha_2)^{\T} \simeq (1-\alpha_1\alpha_2)^{\T -1}$ and $[1+ \alpha_1 \alpha_2 (\alpha_1 - 2)]^{\T} \simeq [1+ \alpha_1 \alpha_2 (\alpha_1 - 2)]^{\T -1}$, and thus the equations above are approximately valid also for computing the average number of stabilized connections of a non-coding neuron and its variance.

\section{Calculation of the mean value of $k'_t$
over the rewiring steps}
\label{app:avg_k1t}
In Section \ref{subsec:rewiring} we obtained the expression of $\SII$ in the presence of rewiring and we observed that this depends on the parameter $k'_t$, given by (Equation \eqref{eq:avg_k1t}):

\begin{equation}
\label{eq:avg_k1t1}
\langle k'_t \rangle = p_t \C ( 1 - \alpha_1 ) ,
\end{equation}

 where, according to Eq. \eqref{eq:pt}, $p_t$ is given by

\begin{equation}
p_t = 1 - (1 - \alpha_1 \alpha_2)^t .
\end{equation}

In order to calculate the mean value of $\SII$ for all patterns, $k'_t$ should be averaged over all the values of the training index $t$ for which the rewiring is performed, i.e.,

\begin{equation}
\label{eq:t_rewiring2}
t = r i \qquad i = 0, \dots , \frac{\T}{r} ,
\end{equation}

where $r$ is the rewiring step and for simplicity we assume that $\T$ is a multiple of $r$ and that there is a final rewiring after the last training step. The average of $p_t$ over the rewiring values of $t$ is

\begin{equation}
\langle p_t \rangle =
\frac{\sum_{i=0}^{\T/r} 1 - [(1 - \alpha_1 \alpha_2)^r]^i }
{\frac{\T}{r} + 1} =
1 - \frac{\sum_{i=0}^{\T/r} [(1 - \alpha_1 \alpha_2)^r]^i }
{\frac{\T}{r} + 1}
= 1 - \frac{b r}{\T + r} ,
\end{equation}

where we introduced a parameter $b$ defined as

\begin{equation}
b = \sum_{i=0}^{\T/r} [(1 - \alpha_1 \alpha_2)^r]^i
= \frac{1 - [(1 - \alpha_1 \alpha_2)^r]^{\T/r + 1}}
{1 - (1 - \alpha_1 \alpha_2)^r}
= \frac{1 - (1 - \alpha_1 \alpha_2)^{\T + r}}
{1 - (1 - \alpha_1 \alpha_2)^r} .
\end{equation}

\section{Noise addition during test phase}
\label{app:noise}
As anticipated in Section \ref{subsec:noise}, in order to assess the generalization capacity of the model proposed in this work, the test input patterns were generated starting from the corresponding training input patterns by adding noise with a given probability distribution.
More specifically, each test pattern is generated by adding to the rate of the corresponding training pattern the contribution of a further extraction from
a truncated Gaussian distribution
$G(\eta)_{\mu_{\text{T}}, \sigma_{\text{T}}}$.
 Therefore the single neuron rate in a test pattern will be given by the following formula:

\begin{equation}
    \nu_{\text{tot}} = \nu + \eta ,
\end{equation}

where $\eta$ is a rate driven by the distribution $G(\eta)_{\mu_{\text{T}}, \sigma_{\text{T}}}$.
The input signal to a neuron of the population $\popII$ can be expressed as the scalar product between the vector 
$\vec{\mathcal{W}}$ of the weights and the vector
$\vec{\nu}_{\text{tot}}$ of the rates of the presynaptic neurons:

\begin{equation}
    \vec{\mathcal{W}}\cdot\vec{\nu}_{\text{tot}} = \vec{\mathcal{W}}\cdot\vec{\nu} + \vec{\mathcal{W}}\cdot\vec{\eta} .
\end{equation}
Since the noise distribution has zero mean, 
its contribution to the average values of the signals
in input to the coding and background neurons,
$\langle\SII\rangle$ and $\langle\Sb\rangle$, will be zero.
On the other hand, it will affect the variance
of the background signal, $\varSb$.
Since $\nu$ and $\eta$ are independent and random variables, the overall variance will be equal to the sum of the variance in the absence of noise $\varSb$ 
(see Equation \eqref{eq:var_sb2_cont})
plus the variance due to noise.
Thus
\begin{equation}
\label{eq:noisy_varSb}
    \sigma^{*2}_{\text{b}} =
    \varSb +
    \langle k \rangle
  (\Wc \sigma_{\eta})^2
  + (C - \langle k \rangle)
    (\Wb \sigma_{\eta})^2 =
    \varSb + \sigma^2_{\eta} \C \Bigl[ p \Wc^2 + (1-p)\Wb^2 \Bigr] ,
\end{equation}
where
$\sigma^2_{\eta} = \sigma^2_{\text{T}}$
is the variance of $G(\eta)_{\mu_{\text{T}}, \sigma_{\text{T}}}$.
Truncating the Gaussian distribution in the symmetric interval $[-2\sigma, 2\sigma]$, the mean is zero, whereas the variance is

\begin{equation}
    \sigma^{2}_{\text{T}}= \sigma^2 \Bigl[1- \frac{4\cdot e^{-2}}{\sqrt{2\pi}\text{erf}(\sqrt{2})} \Bigr] .
\end{equation}

\begin{figure}[H]
    \centering
    \includegraphics[trim=0cm 17.75cm 0cm 0cm, clip=true, width=\columnwidth]{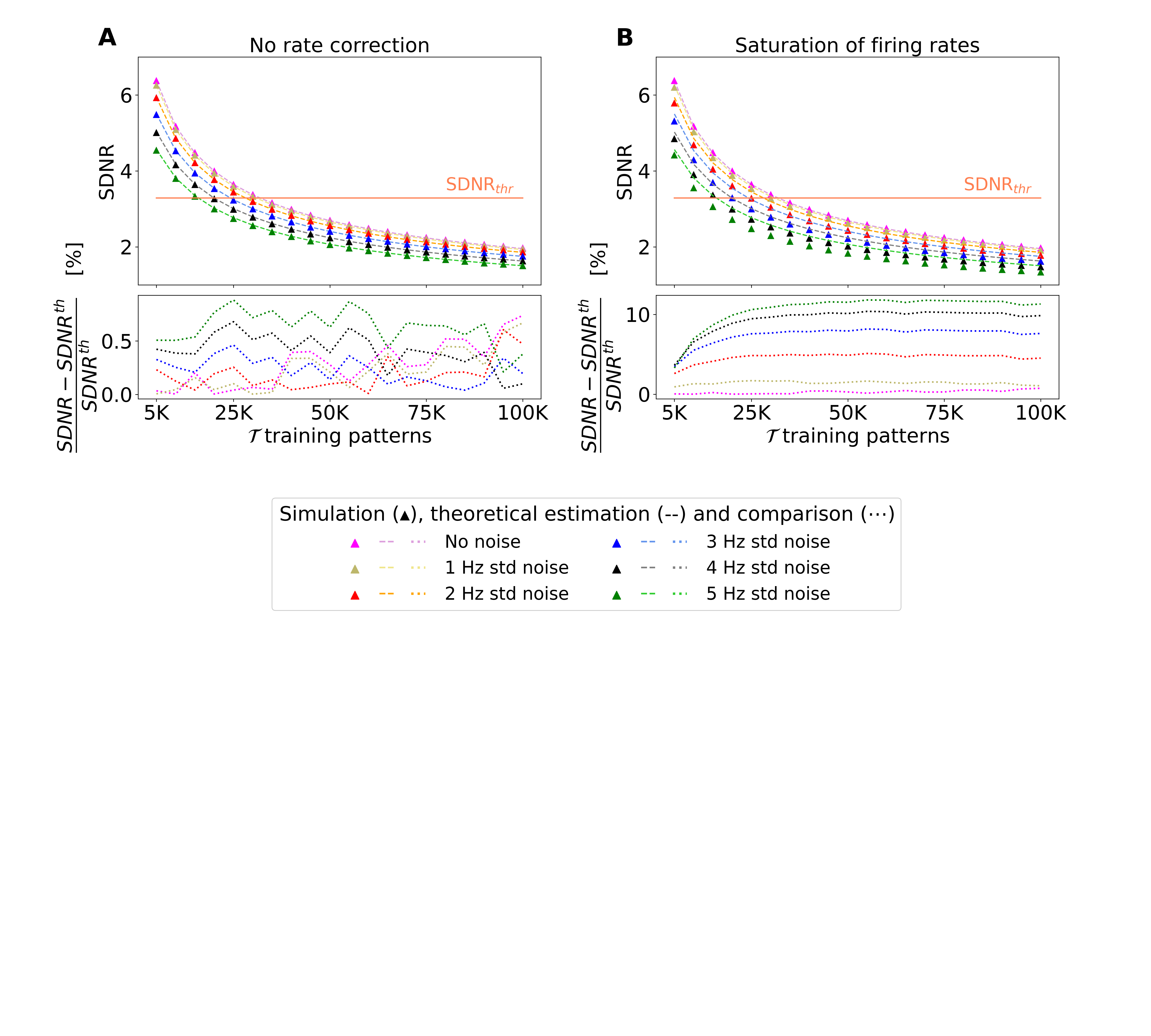}
    \caption{Values SDNR and percent error with respect to the theoretical predictions, as a function of the number of training patterns $\T$ when firing rates values during test phase are not corrected if negative \textbf{(A)} and when they are saturated to zero \textbf{(B)}.
    The different color families identify the simulation and theory results when no noise is provided (magenta-pink), or having a noise standard deviation of $1$\,Hz (dark khaki-khaki), $2$\,Hz (red-orange), $3$\,Hz (blue-light blue), $4$\,Hz (black-grey) and $5$\,Hz (green-light green). The orange horizontal line represents the minimum SDNR for the network to recall the patterns during test correctly.}
    \label{fig:SDNR_comparison}
\end{figure}

As mentioned in the text, adding noise with fluctuations greater than or comparable to the average firing rate can produce negative rate values for a fraction of the neurons. Since negative rate values are not physically possible, this behavior can be corrected in the simulations by simply  replacing negative values of the firing rates with zero, i.e. saturating negative rates to zero. In Figure \ref{fig:relative-error-saturation}, we noticed that when the standard deviation of the noise is equal to $\langle \rl \rangle = 2$\,Hz the differences between simulation and theoretical framework increase up to $5\%$ for the SDNR and more than $10\%$ for $\langle \SII \rangle$, $\langle\Sb\rangle$ and $\varSb$, meaning that the theoretical framework is currently not able to take this effect into account. In particular, 
the SDNR resulting from simulations is smaller than the theoretical estimation. For better visibility, we show in Figure \ref{fig:SDNR_comparison} the comparison of the SDNR with non-saturated and saturated negative rates, respectively, using values of noise standard deviation up to $5$\,Hz.

As can be seen, while the addition of noise without negative rate correction leads to discrepancies between simulations and theory in the order of $0.5\%$ in the case of a noise standard deviation of $5$\,Hz, the same discrepancy with negative rates saturated to zero is around $10\%$. The reason behind this discrepancy is related to the fraction of neurons of $\popI$ that, because of the noise, can result having a negative firing rate. Indeed, the firing rate distribution of the neurons is lognormal with an average near to $\langle\rl\rangle$, so in case of addition of noise with standard deviation compatible or even greater than this value a large fraction of the neurons can have their firing rate set to zero and thus do not project any input to neurons of $\popII$ when saturation is enabled. Nevertheless, it should be considered that the noise levels shown in Figure \ref{fig:SDNR_comparison} are relatively high when compared to the average rate used in these simulations, and thus may lead to significant changes in the rate distributions.
Indeed, a different choice for the values of $\langle \rl \rangle$ and $\langle \rh \rangle$ (and thus a different average rate of the whole distribution) would have an impact on the discrepancies shown here. In particular, a higher average rate would strongly diminish the amount of neurons having a negative firing rate as a result of the noise addition.

\section{\label{app:CsuN} Estimation of the bias on $\varSb$}

\begin{figure}[h]
    \centering
	\includegraphics[width=\columnwidth]{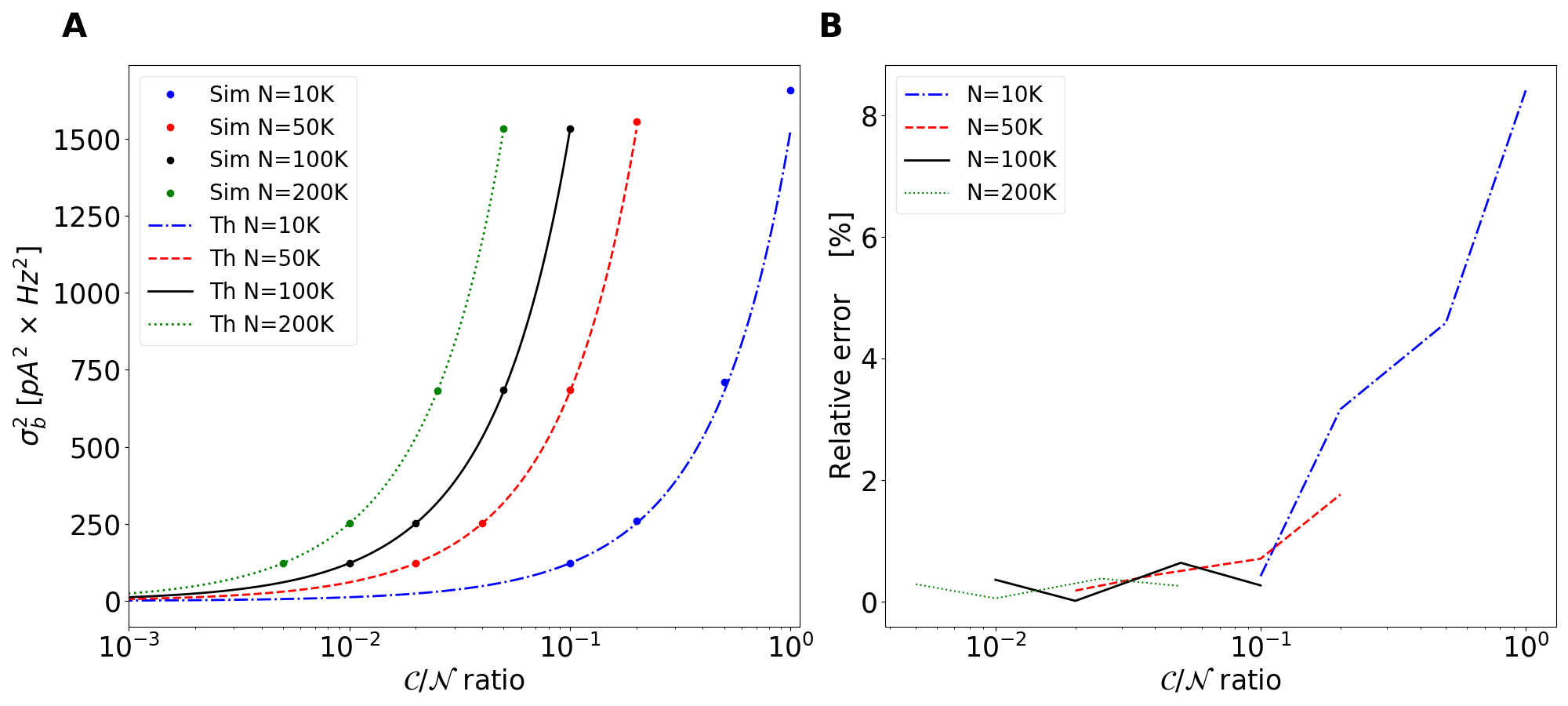}
	\caption{\small 
    \textbf{(A)} Comparison between theoretical and experimental values of $\varSb$ as a function of the $\mathcal{C}/\mathcal{N}$ ratio for different values of $\mathcal{N}$. Lines represent the theoretical prediction (Th), whereas dots represent the values obtained from the simulation (Sim). \textbf{(B)} Relative error between simulation results and theoretical prediction.}
	\label{fig:c_n_comparison}
\end{figure}

As mentioned in the main text, Figure \ref{fig:relative-error} show that the relative error of $\varSb$ is greater than that shown for $\langle \Sb \rangle$ and $\langle \SII \rangle$. This is due to the possibility of input correlation related to random connectivity, which is not taken into account by the theoretical framework. 
In particular, the average number of presynaptic neurons in common to two arbitrary neurons of
$\popII$ depends on the total number of neurons of $\popI$ and on the number of incoming connections per neuron of $\popII$.
Calling $\NI=\mathcal{N}$, we can state that the bias due to this simplification becomes more and more relevant when the ratio $\mathcal{C}/\mathcal{N}$ increases.
In order to estimate this bias as a function of the $\mathcal{C}/\mathcal{N}$ ratio,
we performed a series of simulations with a fixed
number of training patterns, $\T=1000$, changing the $\mathcal{C}/\mathcal{N}$ ratio.
Figure \ref{fig:c_n_comparison} shows the results of this analysis.

As can be seen in the right panel of Figure \ref{fig:c_n_comparison}, a greater value of $\mathcal{C}/\mathcal{N}$ leads to a higher discrepancy between theoretical prediction and simulation. However, such a ratio, for natural density circuits in the brain, is very far from values of $\mathcal{C}/\mathcal{N}$ near unity. Indeed, a plausible value of the ratio would be less than $0.1$, resulting in negligible relative errors.

\section{\label{app:rate-discr} Discrete rate model}
The framework described in this work adopts a general approach for the firing rate distribution when an external stimulus is injected into a neuron population. Indeed, it would be possible to simplify this assumption by considering a discrete distribution, so that neurons can assume, when a stimulus is provided, only two possible values: $\rh$ (high rate) or $\rl$ (low rate). In this appendix, we show that this approach can provide identical results with respect to the most general derivation for a continuous distribution.\\
During training, population $\popI$ is targeted by an external input, so the neuron can have a high or low rate in agreement with the probabilities $\alpha_1$ and $\beta_1 = 1 – \alpha_1$, of falling in the high or low activity regime. The corresponding pattern for the contextual stimulus is generated similarly, extracting the values of the firing rates of the $\popII$ neurons, $\rh$ or $\rl$, with probabilities $\alpha_2$ and
$\beta_2 = 1 – \alpha_2$, respectively.\\
A connection will be stabilized in a training example if both the presynaptic and the postsynaptic neurons assume a high firing rate $\rh$, and the probability that a connection is stabilized in at least one of the $\T$ training examples can be derived at the same way as shown in Section \ref{sec:rate-cont}, so that
\begin{equation}
     p_{\T}= 1-(1-\alpha_1 \alpha_2)^{\T}.
\end{equation}
The average number of stabilized connections is the same as shown in Equation \eqref{eq:av_k}, so
\begin{equation}
     \langle k \rangle=\C \Bigl[ 1 - (1-\alpha_1\alpha_2)^{\T} \Bigr] = \C p_{\T}.
\end{equation}

The test set consists of $V$ firing-rate patterns of the neurons of $\popI$, randomly extracted from the $\T$ input patterns of the train set. In the following, the patterns are unaltered, thus each input pattern of the test set is identical to an input pattern of the train set.
The contextual stimuli are not used in the testing phase.

The average rate of the neurons in the population  $\popI$ is
\begin{equation}
\label{eq:av_r}
    \langle \nu \rangle = \alpha_1\rh + (1-\alpha_1)\rl.
\end{equation}
The input signal targeting a background neuron of $\popII$ is equal to the weighted sum of the signals coming from the $\C$ connections:
\begin{equation}
\label{eq:Sb_discr}
   \Sb = \Wc \sum_{i=1}^{ k}\nu_i +\Wb \sum_{i=1}^{\C- k } \xi_{i},
\end{equation}
where $\C$ is the number of incoming connections, $k$ is the number of stabilized connections, $\nu_i$ are the firing rates of the neurons connected to the stabilized connections and $\xi_i$ are the firing rates of the neurons connected to the unstabilized connections.
From the linearity of $\Sb$ with respect to $\nu_i$ and $\xi_i$ and from the fact that the rates of presynaptic neurons have the same mean value $\langle \nu \rangle$, it follows that
\begin{equation}
\label{eq:av_sb_discr}
    \langle \Sb \rangle = [ \Wc \langle k \rangle 
    +  \Wb (\C - \langle k \rangle) ] \langle \nu \rangle .
\end{equation}

From these results, we can now calculate the variance on the background signal, which is defined as:
\begin{equation}
     \sigma_{b}^{2}=\langle (\Sb-\langle \Sb \rangle)^2 \rangle .
\end{equation}
Using Equations \eqref{eq:Sb_discr} and \eqref{eq:av_sb_discr}, we can compute the variance as:
\begin{equation}\label{eq:discrprelim}
    \sigma_{b}^{2}= \langle \Bigl[ \Wc \sum_{i=1}^{k}\nu_i +\Wb \sum_{i=1}^{\C-k} \xi_i 
    - [ \Wc \langle k \rangle 
    +  \Wb (\C - \langle k \rangle) ] \langle \nu \rangle \Bigr]^2 \rangle .
\end{equation}
Taking advantage of the equality $\langle k \rangle=k+(\langle k \rangle-k)$, we can rewrite:
\begin{equation}
   \Wc \langle k \rangle
    + \Wb (\C - \langle k \rangle)
    = \Wc k + \Wc (\langle k \rangle - k) + \Wb \Bigl[(\C-k) + (k-\langle k \rangle ) \Bigr] =
\end{equation}
\begin{equation*}
     =\Wc k + \Wb (\C- k) + \Wc (\langle k \rangle - k)  + \Wb (k - \langle k \rangle) .
\end{equation*}
Inserting this last expression in Equation \eqref{eq:discrprelim} and rewriting the terms with the multiplicative factors $k$ and $\C - k$ with summations, such as for example $\Wc k \langle \nu \rangle = \Wc \sum_{i=1}^{k} \langle \nu \rangle$, we obtain:
\begin{equation}
\label{eq:var_sb2}
     \sigma_{b}^2 = \langle \Bigl[ \Wc \sum_{i=1}^k (\nu_i - \langle \nu \rangle) +\Wb \sum_{i=1}^ {\C-k} (\xi_i - \langle \nu \rangle) + (k-\langle k \rangle)(\Wc-\Wb)\langle \nu \rangle \Bigr]^2 \rangle .
\end{equation}
Taking into account that the mixed terms go to zero since $\sum_{i} \langle(x_i - \langle x \rangle) \rangle=0$, setting $\sum_{i} (x_i - \langle x \rangle)^ 2=\sigma^{2}_{x}$, we will have that:
\begin{equation}
\label{variance_sigmab}
     \sigma_{b}^2= \Bigl[\Wc^2 \langle k \rangle + \Wb^2 (\C - \langle k \rangle)\Bigr] \sigma_{\nu}^2 + (\Wc - \Wb)^2 \sigma_{k}^2 \langle \nu \rangle ^2 ,
\end{equation}
where $\sigma_{k}^2 = \langle (k- \langle k\rangle )^2 \rangle$.
In the previous formula we note two contributions depending respectively on the variance of the firing rate and on the variance of the number of stabilized connections. The value of the variance of $k$ is derived in Appendix \ref{app:var_k}, whereas the variance of the rate is $\sigma^2_{\nu}=\langle\nu^2\rangle - \langle\nu\rangle^2$. Please note that the variance of the rate differs from the one adopted in the case of continuous rate distribution.\\
To estimate the average input to a coding neuron of $\popII$, we can follow the same derivation shown in Section \ref{sec:rate-cont}, so that:
\begin{equation}
\label{eq:s2_discr}
\begin{split}
    \langle \SII \rangle &= \Wc \C \alpha_1 \rh + \Wc \langle k' \rangle \rl + \Wb (\C' - \langle k' \rangle) \rl =\\
    &= \Wc \C \alpha_1 \rh + \Wc \langle k \rangle (1-\alpha_1) \rl + \Wb (\C - \langle k \rangle) (1-\alpha_1) \rl =\\
    &= \Wc \C \alpha_1 \rh + \Bigl[ (\Wc - \Wb) \langle k \rangle + \C \Wb  \Bigr] (1-\alpha_1)\rl .
\end{split}
\end{equation}

\begin{figure}[H]
\centering
\includegraphics[width=\columnwidth]{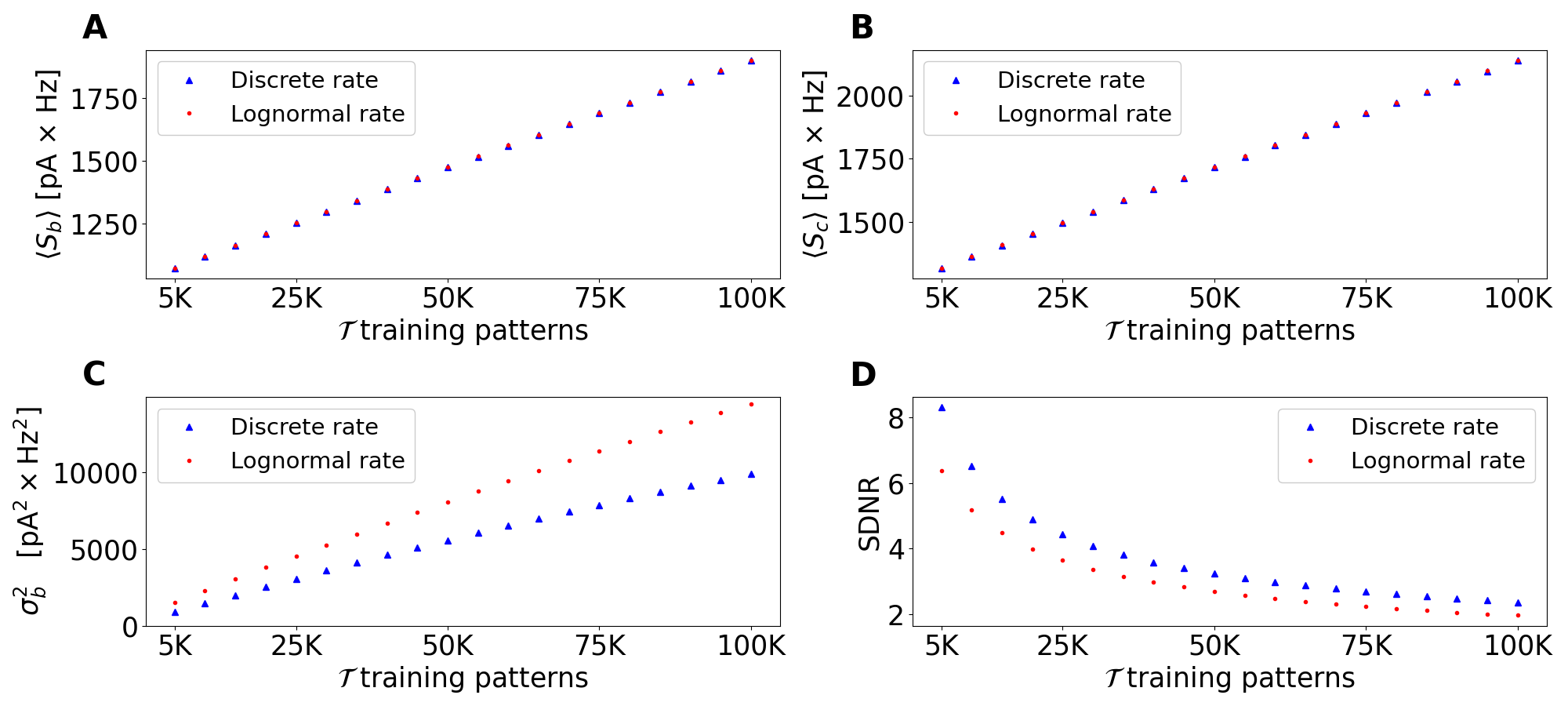}
	\caption{\small 
	Comparison between $\langle\Sb\rangle$ \textbf{(A)}, $\langle\SII\rangle$ \textbf{(B)}, $\varSb$ \textbf{(C)} and SDNR \textbf{(D)} obtained from simulations using discrete (blue line) or lognormal (red dotted line) firing rate distribution. The values are given as a function of the number of training patterns $\T$, and no noise is applied during test. The lognormal rate simulation results are the same as the ones of Figure \ref{fig:relative-error} labeled as "no noise".
	\label{fig:discr_vs_contin}}
\end{figure}

The previous formula does not consider the rewiring of the connections; the effect of rewiring is taken into account in Equation \eqref{eq:s2_cont_rewiring}. In the discrete rate approximation, the equation changes so that instead of having the average high or low rate we simply have the values $\rh$ and $\rl$. Now, it would be possible to obtain the signal-difference-to-noise-ratio (SDNR) using Equation \eqref{eq:SDNR}.\\
Having derived the values of $\langle\Sb\rangle$, $\varSb$ and $\langle\SII\rangle$ in case of lognormal and discrete firing rate distribution, it would be interesting to compare the results of the simulations employing the two approaches. As we discussed, the main difference in calculating $\langle\Sb\rangle$ and $\langle\SII\rangle$ with a continuous rate model versus the discrete model is that the discrete values of $\nu_l$ and $\nu_h$ are replaced, respectively, by the average values of the rate below and above threshold, calculated on the continuous probability distribution. On the other hand, the variance of the background signal differs in the two models, because it depends on the variance of the rate, $\sigma^2_{\nu}$, which depends on the firing rate distribution adopted.\\
Figure \ref{fig:discr_vs_contin} shows the comparison of the simulation outcomes using discrete and lognormal rate values, using the same parameters needed to produce the data shown in Figure \eqref{fig:relative-error}.

We can see that the curves of $\langle\Sb\rangle$ and $\langle\SII\rangle$ obtained from the simulations using the continuous firing rate distribution are superimposed
on those obtained using the discrete model; this is because the choice of the threshold on the lognormal distribution is done so that the average values for low and high rate, $\langle \nu_l \rangle$ and 
$\langle \nu_h \rangle$, correspond to the values adopted in the discrete rate model.
On the other hand, the variance of the background signal $\varSb$ differs in the two models, because it depends on the variance of the firing rate, $\sigma_\nu$, which is different in the two cases. This leads also to the different behavior of the SDNR.

\end{document}